**Could AI Leapfrog the Web? Evidence from Teachers in Sierra Leone**

**Authors:** Daniel Björkegren[1]*, Jun Ho Choi[1], Divya Panchaksharappa Budihal[1], Dominic Sobhani[2], Oliver Garrod[3], Paul Atherton[3]

**Affiliations:**

[1]School of International and Public Affairs, Columbia University; New York, United States of America.

[2]Columbia University; New York, United States of America.

[3]Fab Inc.; London, United Kingdom.

*Corresponding author. Email: dan@bjorkegren.com

**Abstract:** Only 37% of sub-Saharan Africans use the internet, and those who do seldom rely on traditional web search. A major reason is that bandwidth is scarce and costly. We study whether an AI-powered WhatsApp chatbot can bridge this gap by analyzing 40,350 queries submitted by 529 Sierra Leonean teachers over 17 months. Each month, more teachers relied on AI than web search for teaching assistance. We compare the AI responses to the top results from google.com.sl, which mostly returns web pages formatted for foreign users: just 2% of pages originate in-country. Also, each web page consumes 3,107 times more bandwidth than an AI response on average. As a result, querying AI through WhatsApp is 98% less expensive than loading a web page, even including AI compute costs. In blinded evaluations, an independent sample of teachers rate AI responses as more relevant, helpful, and correct answers to queries than web search results. These findings suggest that AI can provide cost-effective access to information in low-connectivity environments.

## 1. Introduction

Traditional web search has been adopted surprisingly little in sub-Saharan Africa. After tremendous investments in network infrastructure by firms, governments, and multilaterals, mobile phone penetration is high (92 phones per 100 people), and 85% of the population lives within access of mobile internet signal (*1*). Yet only 37% of the population uses the internet (*2*). Even more striking is *how* this minority uses the internet: while 76% of internet users across five African countries used social networking daily in 2022, only 29% used web search daily, as shown in Figure 1 Panel A (*3*). This study focuses on teachers in Sierra Leone, knowledge workers who work in remote areas. In our sample that was given access to AI, 91% of teachers report using WhatsApp daily—but only 3% report using web search daily for teaching activities, as shown in Fig. 1 Panel B. Connectivity can improve economic outcomes in low-income countries (*4*, *5*), but this lack of adoption suggests that the full potential of the information revolution has yet to be realized.

A large problem is that bandwidth is scarce and costly. In the region, internet connections are typically pay-as-you-go: users pay for each megabyte or gigabyte of data. In 5 African countries, 52% of internet users report that their usage is limited by the cost of data, and 8% report their usage is limited by network performance or availability (*3*). Most users access the web through mobile devices: across 39 African countries, just 18% of adults report owning a computer (*6*). But traditional web search surfaces prewritten web pages that are formatted for foreign audiences, who have faster internet connections and often use larger computer screens. An audit study found that in sub-Saharan Africa fewer than 30% of a sample of web search results for capital cities contained local results, compared to over 90% in North America (*7*).

We consider how these barriers affect a group of knowledge workers who require reliable access to information. We consider how teachers in Sierra Leone use a new AI chatbot embedded within the communication medium they already use—WhatsApp. The chatbot allows teachers to submit queries to an AI system as they would to a friend or family member. After being trained, 45% of teachers given access to AI use it for teaching activities at least once a month on average, while only 25% report using web search monthly, as shown in Fig. 1 Panel B.

This study considers whether this new technology might lead to a form of 'leapfrogging', where a population that makes little use of an early generation of technology (browser-based web search) directly adopts a later generation (an AI chatbot that is trained on similar underlying web content) (*8–10*). We find that the ideas contained in AI responses are similar to those in pages that are available on the web.[1] However, for the traditional form of the web, the publisher determines the precise contents and format of a piece of content, while AI directly answers the user's query in an appropriate format. In this context, AI makes it easier to deliver concise and useful text content through a low-bandwidth interface that teachers already use. Our study thus builds on a literature that shows that text messages and other text-based interfaces can improve access even when traditional web browsing is impractical (*11–14*). We extend this line of research by studying a modern AI model deployed through a familiar interface.

We study a particular deployment model: a large language model accessed via a WhatsApp chatbot. Our contribution is to describe how 529 teachers use this service over 17 months based on the 40,350 queries they submit, and to compare it to the status quo option for information access. We compare bandwidth and cost for observed queries under the chatbot to the pages that would have been surfaced by browser-based search for the same queries, and assess how independent teachers rate the resulting answers.

[1] That AI surfaces content similar to web pages is not surprising, since web content forms a major part of the training of AI large language models.

This study has two parts.

The first part of the study analyzes how teachers use AI by analyzing their queries. Many teachers continue to use the service over time; after 17 months, approximately 19% of teachers in the earliest cohorts are active per month. 40% of respondents reported sharing their phones with other teachers to allow them to use the chatbot, suggesting they found it valuable.

We also classify the content of queries using a language model. Most queries represent requests for facts or concepts (69%)—queries that one might think would be well suited to web search. Other large categories represent requests for lesson planning and assessment (11%), writing support (5%), and professional development and classroom management (2%). The topical subjects of queries span the curriculum.

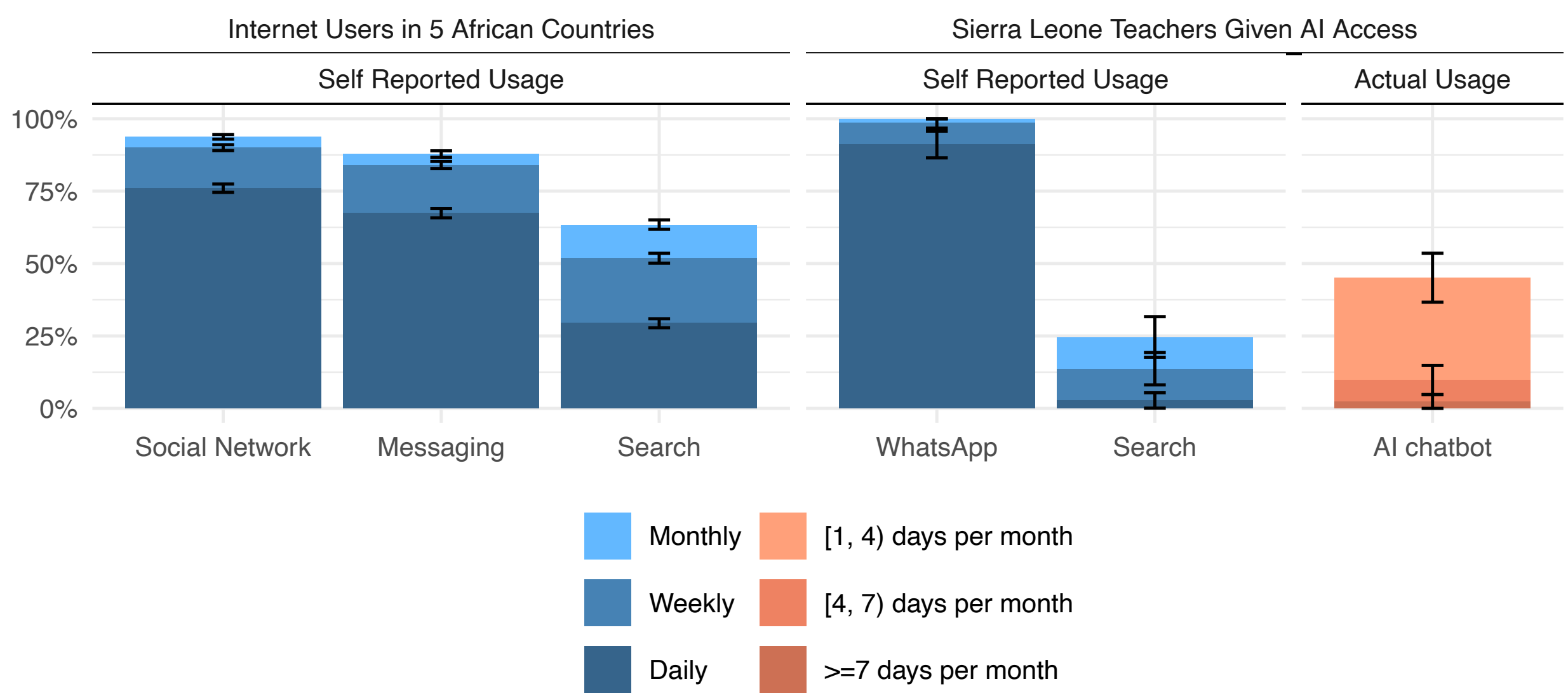

**Fig. 1. Use of internet services in sub-Saharan Africa.** The left panel shows self-reported usage among individuals who used the internet in the past three months, from a representative survey of households in Ethiopia, Kenya, Nigeria, South Africa, and Uganda in 2022 (*3*). The right panel shows usage in the sample of surveyed teachers in Sierra Leone. This includes endline self-reported usage of WhatsApp for texting, and web search for support in teaching activities. The final bar shows actual usage of the AI chatbot, in terms of average number of days with a query submitted per 30 days (month). Error bars indicate 95% confidence intervals for that cumulative proportion. See Figures S1 and S2 for other measures of usage.

The second part of the study compares the responses of AI to the content available on the web, focusing on web pages surfaced by the status quo information retrieval method, web search. We submit a random subsample of teachers' queries to the local version of Google (google.com.sl), which is estimated to have 97% of search engine market share in the country (*15*). We scrape the top 5 search results. Few of these search results are local: only 2.2% of returned web pages are from a Sierra Leonean domain (.sl), and few are from similar countries at all: the vast majority (86%) are from generic or U.S. domains.

We next assess bandwidth usage and cost. We assess bandwidth by loading each web page in a browser that mimics a mobile device, and monitoring network transmissions. The web pages referenced in the top search results consume large amounts of bandwidth, including stylesheets, code, images, and ads. Loading an average web page requires 2,499 kilobytes (KB) of bandwidth, relative to 0.8 kilobytes for the average AI response. Thus, on average a web page requires 3,107 times more data to load than an AI response. Over the connections most teachers have, such data transfers are not only slow, they are costly. Given an estimated mobile bandwidth cost of $1.01 per gigabyte in Sierra Leone (*16*), the web would have cost $2.58 per thousand pages loaded during this period. In early 2022, corresponding AI responses would have cost $30.70 per thousand results, including both the cost of bandwidth and compute. However, the cost of AI compute has declined precipitously, to $2.38 per thousand results in 2023 and to $0.06 per thousand results in 2025. As a result, obtaining the result to an AI query is already 98% less expensive than loading a single web page under the browser conditions we study. Other browsing technologies can reduce bandwidth usage of web pages by loading only selected components like text, but those do not work with every web page and in simulations we find that even partially loading web pages remains more costly than querying AI.

We additionally compare the content of AI responses and the main text of the web pages retrieved by search using a language model. For 74% of queries there is a top 5 search result that contains most of the core ideas of the AI response. Additionally, we rate 48% of queries as being conducive to a single correct answer. These suggest that teachers are primarily using AI to retrieve existing information, rather than create altogether new content.

Finally, we assess the quality of responses by recruiting 25 teachers from around the world to evaluate them. Respondents were shown a sequence of 10 randomly selected queries, along with a response that was randomly selected to be either the AI response or the main text of a top 5 search result for the same query. To avoid bias, respondents were blinded as to the origin of the responses. On a five-point scale, respondents rate AI responses as more relevant (+1.23, p-value < 0.001) and helpful (+1.12, p-value < 0.001). Although it is well known that AI systems may produce incorrect answers (‘hallucinate’), respondents actually rate fewer AI responses as containing inaccurate information (10% vs 31%, p-value of difference=0.023), and on average containing more knowledgeable, correct information (difference +0.6, p-value < 0.001, on a 5-point scale). This appears to be due to two factors. First, teachers’ queries tend to ask for basic information, for which the AI model is more likely to be accurate. Second, respondents’ descriptions of their ratings suggest that individual search results often fall short when viewed as answers to queries. In practice, engaging with search requires synthesizing information from multiple imperfect sources. That modality is suitable on a large screen with a fast internet connection, but is not well suited to our teachers operating on small screens with slow connections. While these ratings capture perceived quality, we do not observe how teachers subsequently use this information in classrooms or its effects on student learning, and we discuss these limitations when considering any implications for policy.

Our findings suggest that content on the web is poorly formatted for low-resource settings, but AI offers promise as a tool to tailor information to the needs and network connections of diverse communities. We demonstrate how AI can tailor content to be accessible in a medium that is well used in our population of teachers, text via WhatsApp. We conclude by discussing open questions about how to best design information systems for low resource settings that encourage critical engagement and represent local needs, particularly for populations speaking languages other than English.

### *1.1 Related Literature*

Much policy around digital connectivity in developing countries has focused on providing access to network connections. However, it has become clear that economic transformation requires more than just access (*17*). This paper joins a literature that considers how individuals in developing countries adopt and use information technologies (*18*–*20*). In particular, it joins studies finding that data costs can be an important constraint on low income people's access to information (*21*, *22*), and that data-efficient text-based services can be effective in such settings (*11*–*14*, *23*). Previous text-based services have broadly fallen into two types. Some allow for open ended questions that are answered by human providers, but these require users to wait for responses and are costly to operate. Alternately, some traditional text services offer instant automated responses, but only to a limited set of commands, and typically surface only select prewritten content. Generative AI enables a new paradigm, allowing for automated question answering services that respond to open-ended queries quickly, at low cost, at scale. It holds the possibility of making information retrieval tools that are useful for broader populations and tasks.

Other work has explored ways to reduce the bandwidth cost of web browsing itself (for example via compression or text-only modes); we focus here on the comparison between the dominant browser-based search experience and an AI chatbot accessed through WhatsApp, which is what our teachers actually use.

We join an emerging set of studies that considers the use of generative AI technologies in real world settings. Otis et al. (*24*) gives entrepreneurs in Kenya two months of access to an AI chatbot that provides business advice, but find that its impacts on business performance are unclear. A larger number of studies analyze impacts of access to AI tools in high-income countries, in either deployments in the field (*25*) or on controlled tasks (*26*–*28*). Experimental studies are most useful when a technology's effects can be summarized by a small number of well-defined metrics. AI, however—much like the internet—is a general-purpose technology which is likely to impact many margins. We study organic usage of AI over 17 months in a population that is underserved by existing information technologies, to understand the gaps in those technologies. Tracking real-world usage patterns maps the channels through which AI is likely to shape behavior and welfare—insights that experimental designs alone struggle to fully uncover.

Relatively high rates of chatbot usage in low-income countries is not unique to our study. A study of web traffic finds that lower middle and lower income countries account for only approximately 15% of search traffic, but approximately 25% of AI chatbot web traffic (*29*).

We also contribute to a literature that considers how individuals use web search. Spatharioti et al. (*30*) analyzes how individuals in a high-income country combine the use of AI and search in a controlled experiment. Ballatore et al. (*7*) analyses the extent to which web search yields local results in different countries, based on queries for the names of a set of worldwide capital cities. But what matters is the performance of these systems on users' actual information needs. Abebe et al. (*31*) analyzes health information needs in Africa based on search queries on the website bing.com. The uptake of our tool provides a unique window into information needs, and allows us to analyze the factors underlying the limits of web search in this underserved population.

Methodologically, we develop language model techniques to extract insight from chatbot queries. We develop prompts to categorize individual queries into conversational threads, determine whether the ideas of one text are contained within another, and determine whether a query is likely to have one or multiple correct possible answers, as well as prompts to help with smaller analysis tasks, all of which are available in the Appendix. Other studies have found that large language models are effective for text analysis (*32*).

Basic descriptives on early usage of this system were described in a previous non-archival workshop paper (*33*), which is subsumed by this paper, which uses new data and is expanded in many directions.

## 2. Context

This study focuses on the education sector within Sierra Leone. The country had a GDP per capita of $509.48 in 2021; the average American has an income 25 times larger when this measure is adjusted for purchasing power (PPP).

Teachers are knowledge workers who regularly must develop content (lesson plans, assignments, and assessments), and navigate complex interpersonal scenarios (with students, parents, and colleagues). Sierra Leonean teachers are stretched thin: there are 42 students per primary teacher, and they have few resources for professional development. The median teacher in our survey had only completed 2 continuous professional development trainings within the past 3 years. Only 19% of surveyed teachers hold bachelor's degrees.

Many teachers work in remote areas that have limited infrastructure. In 2020, only 34% of secondary schools had electricity, and just 8% had internet access. These numbers are even lower among primary schools (6% and 1%). But access to voice and text messaging (2G) is more widespread: 86% of schools nationally were within range of basic mobile phone coverage (*34*), and some teachers are within reach of mobile internet signal (3G or above) at least part of the day. In a survey of teachers, 91% report using WhatsApp daily, as shown in the right panel of Figure 1. They are not strangers to technology.

## 3. AI Chatbot

### *3.1 Design*

From the teacher's perspective, the AI chatbot (called 'theTeacher.AI') appears as a contact in WhatsApp, and they may chat with it just as with any other contact. The design of the system is similar to that of a chatbot for Kenyan entrepreneurs developed and analyzed in (*24*), but is instead tailored for education.

On the backend, the chatbot relays the conversation, plus a system message, to the OpenAI GPT API, and returns the result back to the user. During the period under study, the chatbot used GPT 3.5 Turbo, which was a state-of-the-art model at the time of the first deployment; it has since been upgraded to GPT 4o mini. The system message tailors the system to act like a helpful education assistant, to follow best practice pedagogy (guided by the Education Endowment Fund's reports on education evidence (*35*)), to follow safeguards, and to adapt responses to a teacher who has few classroom resources. The chatbot answers only queries related to education due to budget and for quality control. It was accessible in English, the official language of Sierra Leone; 95% of surveyed teachers in our sample report that English is a main language spoken at home. However, the system message tailors responses to use short, simple language that would be readable to someone without English as a first language.

The core chatbot code is available in an open-source repository and the full system message is given in Appendix B1, so that the system can be replicated by others (see https://github.com/AI-for-Education/fabdata-llm).

### *3.2 Recruitment and Onboarding*

Classroom teachers and principals were recruited in several cohorts from government schools and a non-profit school network (EducAid Sierra Leone). Users were given onboarding training which ranged from one to several hours depending on the cohort. Training including an overview of the system and practice asking questions. During onboarding, teachers were given a training brochure that listed possible uses of the system, including for factual questions, generating content, and getting advice (Figure S9a-c). As part of onboarding, users were asked for informed consent for their anonymized data to be analyzed for research. Only users explicitly given access could use the chatbot. We consider only teachers and principals using the system from Sierra Leonean phone numbers. This yields 529 users, of which 86% are classroom teachers and 14% are head teachers (principals). For simplicity, we refer to users of both types as teachers. We omit queries from each user's first day, to omit any practice questions submitted during the training session.

One cohort of teachers in the non-profit schools, which we call the survey sample, was surveyed three times by EducAid. They were surveyed during onboarding ('baseline', in winter 2024), 2-3 months after onboarding ('midline'), and 3-4 months after onboarding ('endline').

### *3.3 Data*

*Queries.* The chatbot backend records each query submitted by a teacher, a pseudonymous identifier for each teacher, a timestamp, and the corresponding response provided by the AI system. In parts of our analysis, we process queries using a large language model (LLM), using model OpenAI GPT-4o-mini (gpt-4o-mini-2024-07-18) and the prompts described in Appendix B.

*Surveys.* Fab worked with EducAid to develop survey questions covering AI awareness, teaching practices, job satisfaction, and demographics. To the endline survey, the Columbia University researchers added questions about web search usage. The survey sample includes N=166 teachers surveyed at baseline, of which N=143 could be linked to the query data. That declines in the midline (N=152 surveyed, N=133 linked) and endline (N=146 surveyed, N=128 linked).

*Online survey experiment.* The Columbia University researchers additionally conducted on online survey experiment, detailed in Section 4.2.

### *3.4 Usage*

Figure 1 Panel B shows suggestive indications of a leapfrogging pattern in the survey sample. Teachers use search infrequently: after being provided and trained with the AI WhatsApp chatbot, only 3% report using web search daily to help with teaching, and only 25% report using search at least monthly. In contrast, 45% use the AI chatbot at least monthly on average. At the weekly and daily level, the proportion of teachers using the technologies are similar. In Figure S2 we report alternate measures, including self-reported AI usage. More teachers report

consulting the AI chatbot for information about teaching (77.4%) than the internet (37.0%); the proportion reporting consulting the chatbot is not statistically significantly different from the proportion consulting books (68.5%). Because of the limits of self-reported usage data, we focus our analysis on direct usage measures.

We break down patterns of AI usage in more detail in Table 1. The median teacher uses the chatbot 0.9 days per month on average and submits 2.4 queries per month; however, the top 10th percentile of teachers uses it 4.5 days and submit 19.8 queries per month.

Additional patterns are shown in Figure 2. Panel A shows a plot of retention, with the proportion of users submitting a query each month after being onboarded onto the system after a given number of months. 62% of onboarded users are active in the first month; this decays over time to roughly 19% active per month by the 17$^{th}$ month. This could result if, for example, some teachers have either obtained the guidance they need, face barriers to access, or are not satisfied with the system's current design. Most usage occurs during school hours, but there is also substantial usage at night, suggesting it is used for planning, as shown in Panel B.

In an endline question, 40% of respondents reported sharing their phones with other teachers to allow them to use the chatbot, suggesting they found it valuable.

*Heterogeneity*

We assess heterogeneity in usage in Fig. 2 Panel C. Differences are not statistically significant, but patterns are suggestive. The small number of teachers with a bachelor's degree use the assistant 110% more per month on average, though this is partly influenced by one high use teacher. This pattern of usage corroborates other work in low income countries that find AI assistance benefits higher performing entrepreneurs more (*24*), but contrasts with several studies in high income countries finding that AI assistance benefits lower skilled workers more (*25*, *28*). Classroom teachers use the chatbot slightly more than head teachers. Women are slightly less likely to use the chatbot than men, echoing other studies (*37*, *38*). Teachers who used WhatsApp more frequently at baseline use the chatbot more. In contrast, it is the teachers who rarely use search at endline who use the chatbot more. This pattern would be consistent with the two technologies being substitutes, or with the chatbot working for teachers for whom search is less useful. (Heterogeneity by job position is computed for the full sample; heterogeneity by other characteristics is computed for the survey sample.)

**Tab. 1. Usage of AI chatbot.** Table shows the distribution of usage by teacher, over the timespan from a teacher's first query to the end of the dataset, with one month computed as 30 days.

| | Mean | Q25 | Q50 | Q75 | Q90 |
|---|---|---|---|---|---|
| Active days per month | 1.7 | 0.2 | 0.9 | 2.4 | 4.5 |
| Queries per month | 7.5 | 0.4 | 2.4 | 9.8 | 19.8 |
| Queries | 76.3 | 4.0 | 24.0 | 90.0 | 208.2 |
| | | | | | |
| **N** | 529 | | | | |

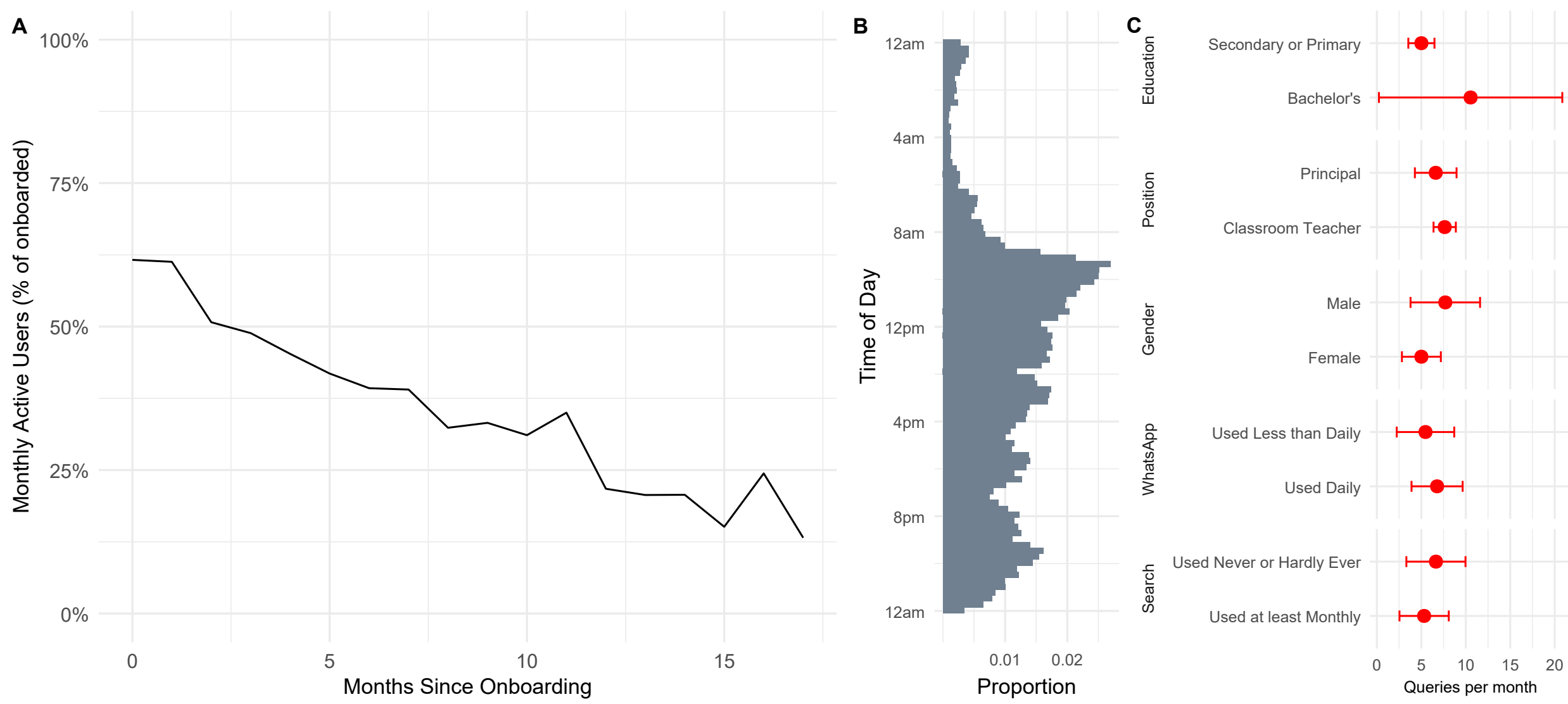

**Fig. 2. Usage of AI chatbot.** Panel (**A**) shows that many teachers continue to use the chatbot, though there is a decline over time. Panel (**B**) shows usage by time of day. Panel (**C**) shows heterogeneity in usage. Position is measured for the entire sample; the remainder are measured in the surveyed sample (WhatsApp use at baseline; education and gender at midline; search at endline). Whiskers display 95% confidence intervals.

*Threads*

Some of the messages sent by teachers continue the existing conversation and some start new conversations. The WhatsApp interface does not have a button to start a new conversation, so we develop an algorithm to divide each user's interactions with the chatbot into separate threads, defined as conversations aimed at a common goal. For each user, we processed the queries and the chatbot's responses sequentially, using a language model. Given a current proposed thread, we assessed whether a subsequent query-response pair should be included in the ongoing thread based on whether the new message continued the conversation, referenced previous messages, requested further clarification, or introduced new information relevant to the ongoing discussion. Queries separated by more than 48 hours were considered to belong to different threads, regardless of the other criteria. If the new message was determined to belong to the currently-examined thread, it was appended to the current thread. Otherwise, the current thread was closed, and the new message was treated as the start of a new thread. This process was repeated for the entire sequence of messages to classify them into distinct threads. The LLM prompt is given in Appendix B2. Based on this classification, 66% of queries start new conversations.

*Tasks and Subjects*

We next categorize the threads into tasks and subjects. We do this in a three step process. We first send the queries to a language model to suggest a list of categories. We manually edit this list to simplify and remove duplicates. We then send each thread to a language model and ask it to classify each one, using the prompt in Appendix B3.

The largest task category is for facts and concept clarification (69%), followed by lesson planning and assessment (11%), writing support (5%), and professional development and classroom management (2%), as shown in Figure 3 Panel A. Thus, the largest category of queries represents retrieval requests that one might have expected search engines to be effective at answering.

We also apply a similar classification algorithm to the subjects of threads. These span the curriculum, as shown in Figure 3 Panel B.

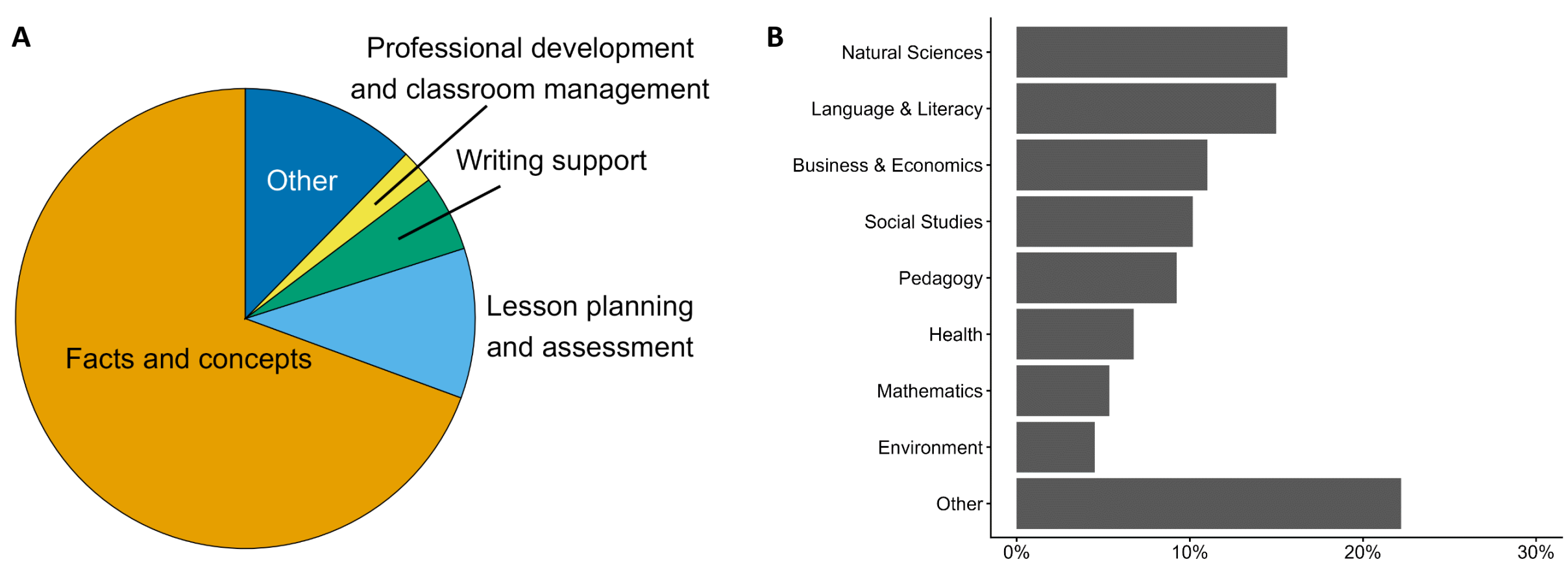

**Fig. 3. Categories of chatbot queries.** Classification of the threads submitted to the chatbot, by the type of task (Panel (**A**)) and subject (Panel (**B**)).

## 4. AI versus Search

We next consider how AI compares to search.

### *4.1 Self-assessment*

The endline survey asked teachers how they determined when to use AI versus search; results are shown in Figure S5. When asked about times they use theTeacher.AI instead of search, all respondents reported at least one reason they did so; however, only 45% reported at least one reason when asked about times they used search instead of theTeacher.AI. The most common reasons to use AI were that chatbot gives a useful answer right away (85%), is more concise (78%), or more trustworthy (72%). 66% of teachers report using AI because it creates content that cannot be found on the web; we quantify this overlap in the next section. 65% report using AI because it works better with the level of internet access. Although 58% of teachers note AI's ability to ask follow ups, this is only occasionally used in our data. The most common reason that a respondent would use search instead of AI is that the web has content that the chatbot does not (18%). As a free response to this question, 5 teachers (3.4%) reported that since receiving access, they have stopped using web search and now instead use theTeacher.AI.

### *4.2 Comparison of AI and search results*

We compare AI and search more systematically by analyzing the queries themselves. We determine the content that teachers would have seen had they submitted the same queries to a search engine rather than AI.[2] We measure the country origin of that content, and compare data efficiency, cost, and quality of the resulting web pages to the AI responses.

We take the queries submitted to the chatbot, and submit them to the local version of Google (google.com.sl), which is estimated to have 97% of search engine market share in the country (*15*). We consider only the first queries asked in each thread to give both technologies similar context, since otherwise AI would respond to natural language follow ups differently than search. We omit queries that are 'fillers', such as introductions or greetings (see Appendix B4). We compute the following statistics for a 12.5% random subsample of queries. Figure S8 shows a diagram of major steps in our analysis pipeline. We report more details and validation exercises in Appendix A.

We use several settings to mimic the search results that would be seen on a mobile phone physically located in Sierra Leone. We queried through internet protocol (IP) addresses in Sierra Leone using a proxy service (Oxylabs). We initialize a web browser with settings that mimic a mobile phone (using Chrome Webdriver, Selenium, and Selenium Wire), emulating the phone model Tecno Spark 8C, which is a common brand in sub-Saharan Africa. We then submit the query to google.com.sl. Once we submit the query, we retrieve the top 5 organic search results, omitting any special sections of the search results page, such as videos, images, and shopping.

While the AI chatbot responds to queries with a custom, direct response, traditional web search requires two steps. First, a search engine presents a customized list of results to the user; then, the user may select one or more preformatted web pages to load. We consider these steps in turn.

In the list of search results returned, very few web pages are local. Only 2.2% of returned web sites are at a Sierra Leonean domain (.sl), and few are from similar countries at all: 86% are generic or largely US hosted (.com, .org, .edu, .net, .gov) and 3% are UK domains, as shown in Figure 4. A breakdown by domain is given in Table S1.

[2] If individuals write queries differently for AI than for search, that could bias our results in favor of AI. However, we find that many of the queries in our data mimic the short, keyword driven style that is common in web search. Further, most of these queries surface web pages that contain most of the ideas in the AI responses (as shown in Figure 7), suggesting that any such bias is likely to be small. Since our population of teachers use web search so rarely, an equivalent dataset of queries from search would be sparse.

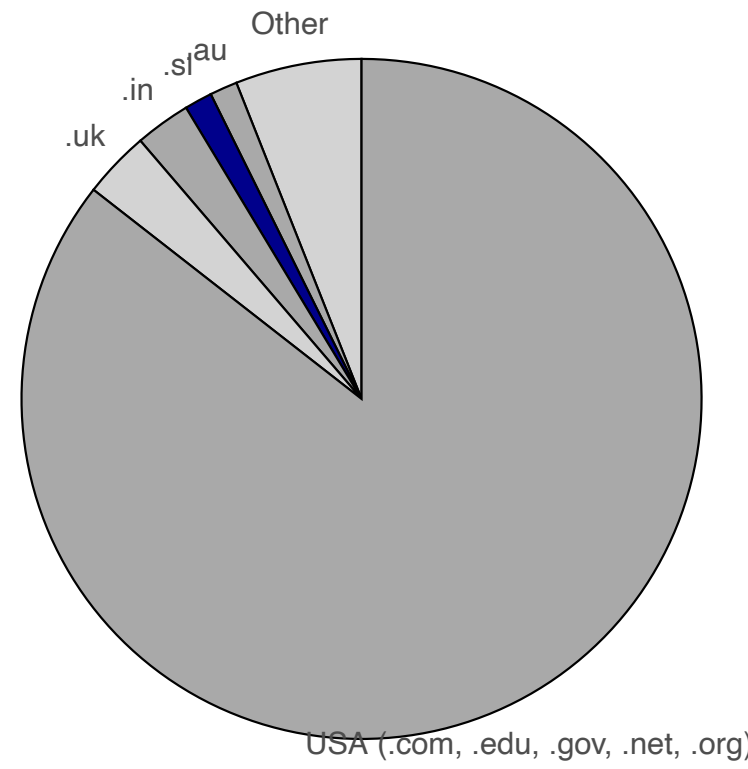

**Fig. 4. Origin domain of search results for Sierra Leonean teacher queries.** The top-level domain (or country code) of the top 5 search results from Google.com.sl for queries submitted to the chatbot. (Sierra Leone domains emphasized in dark blue.)

The main part of our analysis focuses on the second step of search: the web pages that are surfaced. We scrape the main content of the first 5 search results, by downloading each referenced web page and extracting its main text. Standard web pages include text that is not part of the main content (such as navigation links), so for these we extract only the main article using a common automated method to remove extraneous content. For PDFs, we convert the entire file to text. We omit search results that have main content that is video rather than text (the 1.4% of results pointing to YouTube), as well as from a set of domains that commonly block content behind a paywall or login form (4.4% of results). We also omit 17.3% of search results for which we encountered scraping difficulties. Three example queries and responses are given in Tables S2-S4.

For a random sample of 1,000 queries we monitor logs of network transmissions as the page loads to assess the data transfer required for the page and its embedded resources (such as code, styles, fonts, and images). We request that the server compress transmissions to apply any optimizations for low connectivity users. To ensure the page is fully loaded, after the body is loaded, we wait 15 seconds and then add additional waits of 2 seconds up to 5 times if the page continues interacting with the network. Validations suggest our data use measure for search results is conservative. For PDF files we monitor the size of the download. For more details on methods see Appendix A (c) and (d).

*Format*

AI responses are shorter by an order of magnitude (mean word count of 156 versus 3,311). Both because AI responses are shorter, and because they are text (without syntax, styles, images, ads, and code), they require transmitting much less data, as shown in Figure 5. The average data needed to transfer an AI response is 0.8 kilobytes (KB), relative to 2,499 KB for one web page. Thus, a web page requires 3,107 times more data to load than an AI response on average. The data used by an individual web page is in addition to the data needed to transfer for the list of search results in the first place. Larger file sizes not only incur larger transfer costs, they are also challenging to download on slow and intermittent networks.

Overall, relative to chatbot responses, web content is poorly formatted to be viewed on the small devices and low bandwidth connections used by our sample of teachers. Many web pages rely on data heavy elements like JavaScript code, large images, and stylesheets. The web does possess some high-quality local content; for example, 1.6% of search results are from the Sierra Leonean education ministry (representing 74% of all results from Sierra Leone), including localized lesson plans and syllabi. But among those, 92% are portable document files (PDFs) formatted for printing, some over a hundred pages long and difficult to view on a small screen.

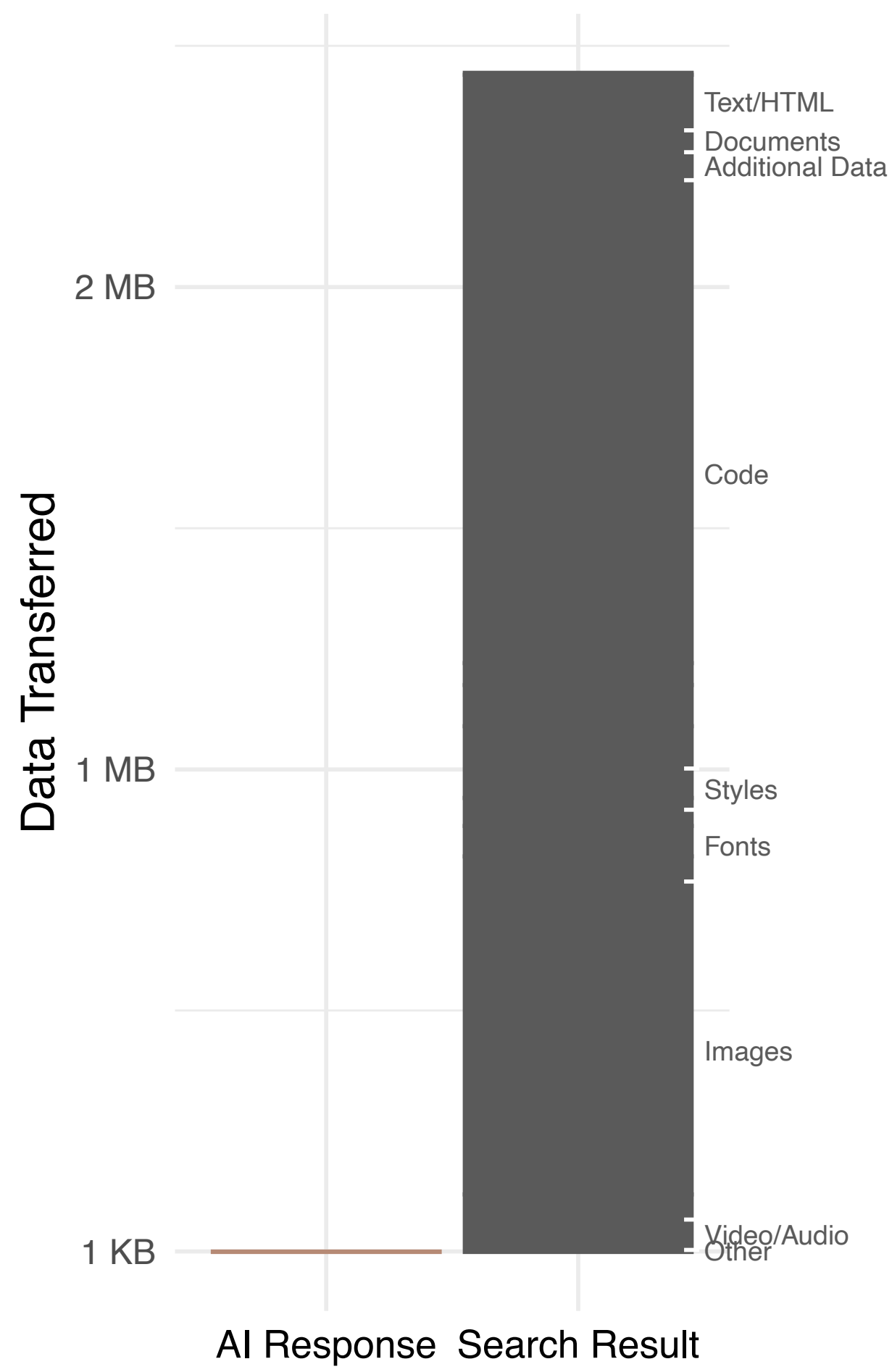

**Fig. 5. Data usage of average AI response and web page.** The average web page surfaced by web search consumes 3,107 times more data to load. The notches on the right break down bandwidth usage by component, including text/HTML, syntax, documents (such as PDFs and Word documents), additional data transmissions, code for rendering and interaction, formatting, images, and media. Content type classified by file extension or MIME type.

There exist browsers that lower bandwidth consumption by attempting to load only select components of a web page (such as omitting images, or loading only the basic text and HTML syntax). However, much of the content of modern websites is essential for rendering the website, including documents (when the main content is a PDF or Word document), JavaScript code, styles, and data transmissions. We break down the data use of the average web page into its components in a set of notches in the right side of Figure 5.

*Cost*

The number one reported limiter that prevents individuals in sub-Saharan African countries from using the internet more is cost (*3*), as shown in Figure S3. Web pages use much more mobile data to download, but AI responses cost more to compute. We compare the costs of the two systems.

For a given query $q$, we compute a lower bound of the cost of search, based on the cost of loading a single web page,

$$c_s(q) = p_{transfer} GB(x_s(q))$$

and compute the total cost of an AI response as

$$c_{AI}(q) = p_{transfer} GB\big(x_{AI}(q)\big) + p_{compute,in} Tokens(S + q) + p_{compute,out} Tokens\big(x_{AI}(q)\big),$$

where $x_j(\cdot)$ represents the content retrieved by method $j$ in response to a query, $GB(\cdot)$ counts the bandwidth required to transmit that content, $Tokens(\cdot)$ computes the number of AI tokens used to represent a given amount of text, and $S$ represents the total tokens used for the system message. Prices include the cost per gigabyte of data transfer ($p_{transfer}$), and the AI compute costs per input and output token ($p_{compute,in}$ and $p_{compute,out}$, respectively). We compute costs for GPT3 Davinci until Nov 2022, GPT3.5 Turbo until July 2024, GPT4o mini until August 2025, and then gpt-oss 20b, using posted prices as displayed in Table S7.[3] For the web, the total cost also represents the cost to users; for AI, users paid only for bandwidth in this implementation.[4]

We plot a cost comparison in Figure 6 for the average web page and the average AI response. Given a mobile network cost of $1.01 per gigabyte in 2024 (*16*), using search on the Sierra Leonean mobile network cost $2.58 per thousand web pages loaded during this period (dashed gray line). AI responses (salmon colored line) cost a total of $30.70 per thousand results in early 2022. However, the cost of AI has declined precipitously, to $2.38 in 2023, to $0.30 in 2024, and to $0.06 per thousand results in 2025. These cost comparisons suggest it is already 42 times more cost effective to query AI than to load a single web page.

[3] We replace a previous model with only lower cost models that are more highly rated on LMArena.com.
[4] Since the beginning of the study, the required subsidy for AI compute fell from about $2.38 to $0.06 per 1,000 queries, so we observe usage in the increasingly common environment in which access to basic AI is cheap or free. Consistent with (36), such a design is better suited to generalizing to future low-cost settings than designs that expose users to the historical high costs.

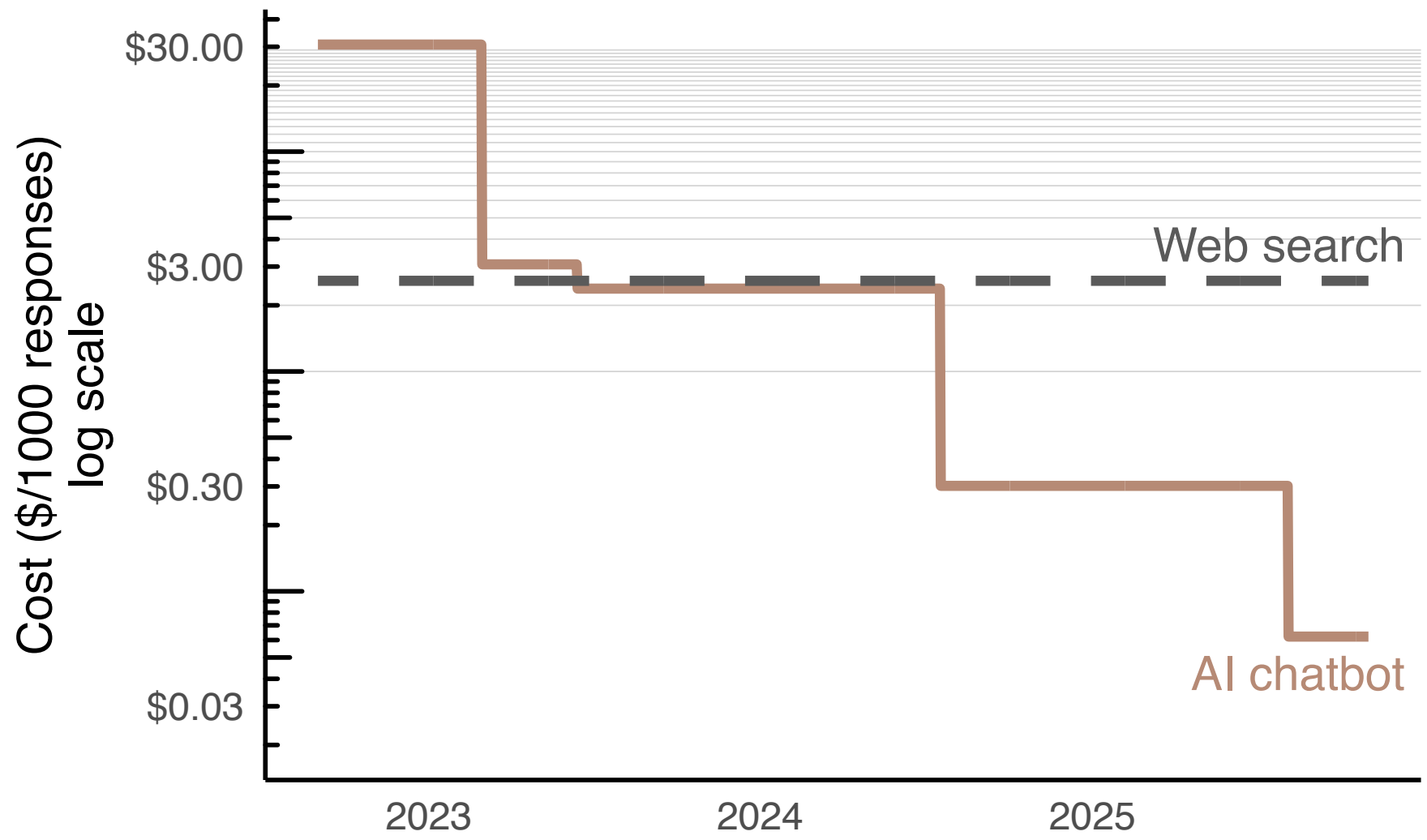

**Fig. 6. Cost comparison over time.** The cost per query is plotted over time, for the average AI chatbot response and search result, in log scale. Costs include mobile data transfer in Sierra Leone of $p_{transfer}$ = $1.01 per gigabyte (*16*), and, for AI, the compute cost for OpenAI GPT models (GPT3 Davinci until Nov 2022, GPT3.5 Turbo until July 2024, GPT4o mini until August 2025, then gpt-oss-20b) applied to the number of tokens used for the query and model output.

We present costs under alternate assumptions in Table S8.

For the web, our main calculation omits the bandwidth required to transmit the search result page itself, and only considers the bandwidth to load a single search result. A search interaction that loads multiple results would cost multiples of the amount quoted ($5.17-$12.90 per thousand queries to load 2-5 web pages). It may be possible to reduce costs by using a low-bandwidth browser that is text-only or otherwise loads only portions of the web page. We simulate the cost of using such browsers by omitting the bandwidth contribution of different components. Omitting images, video, and audio would reduce the cost to $1.77 per thousand pages loaded; omitting all but text/HTML, documents, and code would reduce the cost to $1.47 per thousand pages loaded. If code were also omitted, the cost would be $0.17 per thousand web pages, but many web pages would not function, so one may need to load multiple web pages in order to obtain one that works.[5] Loading multiple results in this mode would again raise the cost (to $0.34-$0.85 per thousand queries for 2-5 pages).

For AI, for comparability our main statistic ($0.06 per thousand responses) considers an AI system without a chat history because we do not factor in follow up queries for either technology. With a completely full chat history, AI's cost would be $0.15 per thousand results. If the steep downward trends in AI computation costs underlying Fig. 6 continue, AI will become even more cost effective. AI's bandwidth cost alone would be $0.001 per thousand results. That cost coincides with the cost users face under the deployed system, which mimics a scenario for the long run cost of such a system if current trends continue.

[5] For example, low-bandwidth browsers like Opera Mini are not fully compatible with JavaScript. As shown in Figure 5, JavaScript code is an important portion of the web pages loaded in these search results.

Search retrieves content from the web and delivers it verbatim, this comparison suggests that in this context it is far more efficient to tailor knowledge on the cloud and deliver only what the user needs.

*Content*

For each query, we then compared the content of the AI response to the main content of the top 5 search results using a large language model (see Appendices B6 and B7 for prompts).

Search results contain similar information as AI responses. We prompt the model to assess the degree to which the text of the search result contains the main ideas in the AI response, on a scale from 1-10, where 1 indicates that the search result has no core ideas of the AI response, and 10 indicates that it captures all core ideas. If the article is longer than 2000 characters, we first summarize it with an LLM. The distribution of contains scores is plotted in Figure 7. We find that 46% of top 5 search results contain most of the core ideas of the AI response (score 6 or above), and for 74% of queries there is a top 5 search result that contains most of the core ideas of the AI response.

We additionally assess whether the query itself is conducive to a single correct answer (like a factual question) or multiple possible correct answers (like asking to generate a story) by asking a language model to classify it. By that classification, 48% of queries are conducive to a single correct answer. We plot the distribution of these two variables together in Figures S6 and S7, showing that contains scores are higher for queries rated as having a single correct answer. Overall, these results suggest users are mostly asking the AI system to retrieve information, and less so to generate new content.

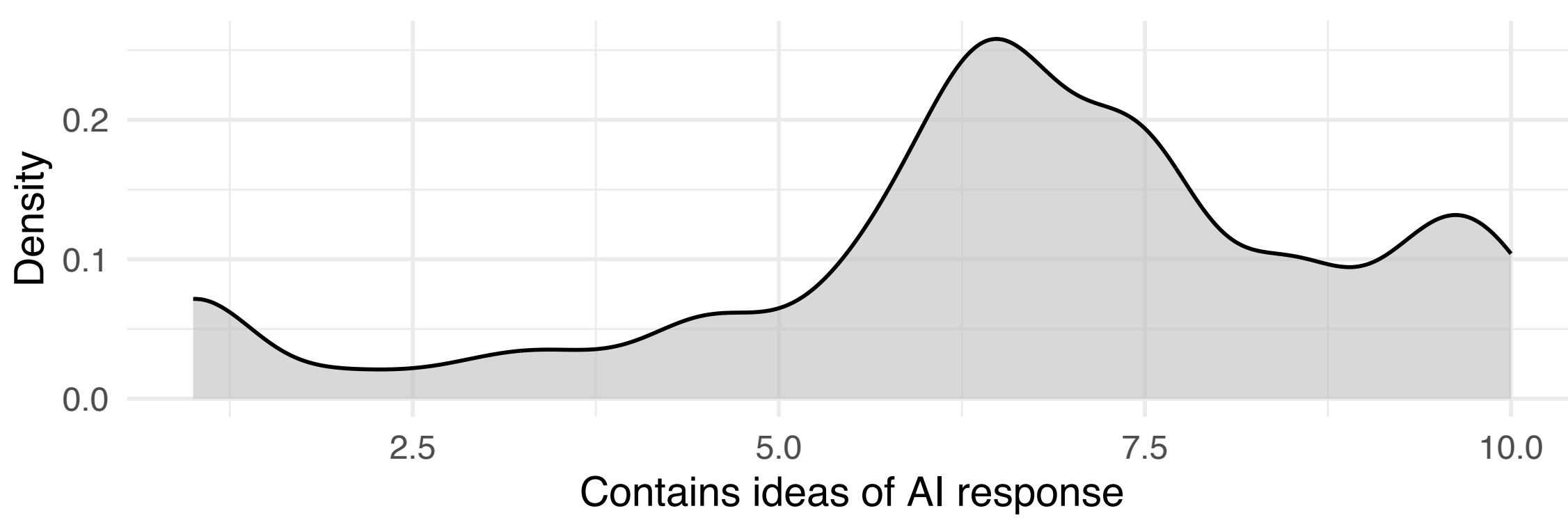

**Fig. 7. Do Teachers use AI to Retrieve Information Already Available on the Web?** Plot shows the distribution of contains scores: a rating of the degree to which the search result contains the main ideas of the AI response for the same query, as assessed by a language model with prompt given in Appendix B6. The search result with the highest contains score for each query is plotted.

*Quality*

We next compare the quality of the text of the responses. We recruit 25 teachers from around the world to evaluate them. We drew a random sample of 100 queries, restricting to those for which both systems produced viewable text (search results successfully scraped and AI system did not refuse to answer). For each, we assembled a dataset with the query, the AI response, and extracted text from one of the top 5 search results (selected at random).

Respondents were recruited on the Prolific survey platform from a pool of teachers. Each was shown a sequence of 10 queries randomly selected from the 100. Alongside each query, they were shown a single response, and asked to rate it along various dimensions. The response was randomly selected to be either the AI response (2/3 probability) or the extracted search result text (1/3 probability). We oversampled AI responses because they are shorter and thus faster to evaluate, and because we wanted to obtain more precise measures of their factuality. To avoid bias, respondents were blinded as to the origin of the responses, both in general, and for each query.[6] They were informed that in rating the response they could use their own knowledge or consult outside resources, such as books or the web, but were asked not to consult an AI chatbot. Respondents were asked to report the resources they consulted. Because we compare only the text, this experiment isolates the difference in content (from AI versus web page) from differences in modality (text-based vs. multimedia web browser). See Appendix A (f) for more details on the structure of the survey.

We present results in Figure 8. On a five-point scale, respondents rate the AI response as more relevant (+1.23, p-value < 0.001) and helpful (+1.12, p-value < 0.001; two-sided F-tests of equality with clustering by query and respondent).

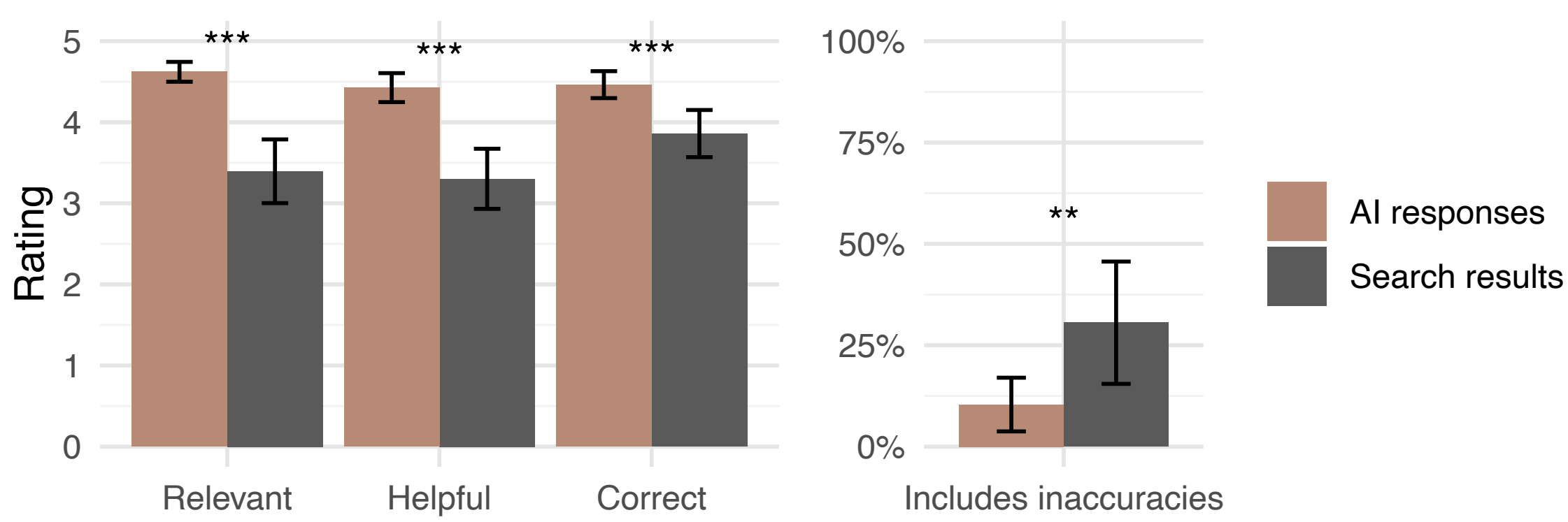

**Fig. 8. Independent teacher ratings of AI responses and web search results.** For a sample of randomly selected queries, with associated AI response and search result randomly drawn from top 5 results. 95% confidence intervals shown with whiskers; standard errors clustered by query

[6] If respondents had known that the potential origins included web pages and AI generated text, they might be able to infer the origin of a given piece of text based on stylistic characteristics. For this reason, the survey never mentioned the possible origins of the texts, nor that the study's aim was to compare the quality of texts from different origins.

and respondent. Stars represent significance level of a two-sided F-test of equality; ***: p<0.01, **: p<0.05.

There is a general concern that AI systems may generate responses that are not correct ('hallucinate'), either not internally consistent with the instructions given, or not factually consistent with the world (*39*, *40*). The random sample of 100 AI responses includes at least three hallucinations that we verify independently. However, respondents rate fewer AI responses as containing inaccurate information (10% vs 31%, p-value of difference=0.023), and on average containing more knowledgeable, correct information (difference +0.6, p-value < 0.001, on a 5-point scale).

Why do respondents rate the search results as having more inaccuracies? Respondents were asked to describe their reasoning in a free response question. Some explain that they considered the text of a web page inaccurate because it made unfounded claims. Other results did not directly respond to the query, included irrelevant information, or did not completely answer the query. For example, a query about 'effect' returned a result about 'affect.' In analyzing these results, it became clear that in the typical search modality, one does not expect individual web pages to necessarily be accurate, or to answer the question, but rather one is expected to compare the veracity and relevance of multiple sources. But that comparison is challenging on small screens and requires even more data than loading a single response. Our experimental structure that blinds the respondents to the source of the response makes clear that individual web pages often fall short when viewed as answers to questions.

We interpret the positive ratings of correctness to suggest that the information teachers are asking for is generally well known by the system, for which hallucinations are less common. Corroborating this, in the endline survey of teachers using the system in Sierra Leone, 72% of teachers reported that when they use AI rather than search a reason is they found AI more trustworthy. Yet it is well known that AI can make mistakes, so an interface that conveys a single definitive answer can be misleading. This is particularly important for users who do not have external resources with which to validate responses.

*Framework for welfare impacts from AI*

To understand the overall welfare impact of AI relative to web search in this context, we describe a simple model of consumer choice. We write a teacher's utility

$$U(Q_s, Q_{AI}, O) = u(Q_s, Q_{AI}) + O$$

as a function of the number of search queries $Q_s$, the number of AI queries $Q_{AI}$, and consumption of all other goods $O$ (denominated in units of currency). We assume $u' > 0$ and $u'' < 0$ in each argument. They face the budget constraint $\bar{c}_s Q_s + \bar{c}_{AI} Q_{AI} + O \leq Y$ given income $Y$ and average cost per query $\bar{c}_s = c_s(\bar{q}_s)$ and $\bar{c}_{AI} = c_{AI}(\bar{q}_{AI})$, for an average query of length $\bar{q}_s$ for search and $\bar{q}_{AI}$ for AI.

Prior to the introduction of AI, the teacher solves

$$\max_{Q_s, O} u(Q_s, 0) + O$$

subject to the budget constraint. Denote the initial optimum as $(Q_s^0, 0, O^0)$.

Once AI becomes available, the teacher chooses $(Q_s^*, Q_{AI}^*, O^*)$. We assume that teachers can evaluate the quality of AI so that this usage is optimal. The welfare impact of introducing AI is

$$\Delta = u(Q_s^*, Q_{AI}^*) + O^* - u(Q_s^0, 0) - O^0.$$

Substituting in the budget constraint and writing $\bar{c}_{AI} Q_{AI}^* = \bar{c}_s Q_{AI}^* - (\bar{c}_s - \bar{c}_{AI}) Q_{AI}^*$, we have

$$\Delta = \underbrace{(\bar{c}_s - \bar{c}_{AI}) Q_{AI}^*}_{Immediate\ cost\ difference} + \underbrace{u(Q_s^*, Q_{AI}^*) - u(Q_s^0, 0) + \bar{c}_s(Q_s^0 - Q_s^* - Q_{AI}^*)}_{Change\ from\ quality\ and\ utilization}.$$

Because the initial allocation $(Q_s^0, 0, O^0)$ is still attainable after the introduction of AI, the model suggests that the overall welfare gain is weakly positive ($\Delta \geq 0$). Part of the utility change is from the cost difference of using AI instead of search for the queries routed to AI, captured in the first callout ('immediate cost difference'). Our empirical results find that teachers use AI and it is much lower cost ($\bar{c}_{AI} \ll \bar{c}_s$), which shows that the first channel is present. However, we find that global teachers rate AI responses as higher quality; the second callout ('change from…') captures the utility gain from this change in quality of information services, and overall changes in utilization which can result from both factors.

Generally, AI is expected to lower the cost of many knowledge tasks. As a result, AI could have positive welfare impacts even in settings where its direct effects on focal outcomes are zero (or negative), because it could free up spending for other uses. Alternately, even in settings where AI has positive direct effects, those effects may come at the cost of increased utilization which takes resources away from other uses. Thus, it may be important to analyze costs when evaluating the effects of AI.

## 5. Discussion

Our results suggest that way web content is currently packaged and delivered contributes to low use of traditional web search in low resource settings, whereas an AI-powered WhatsApp service can meet information needs at much lower cost.

In the traditional web, results are limited to existing content in its existing format. There is little such content from low-resource communities, so these communities rely on content formatted for wealthier regions. In recent years, network connections in wealthy regions have dramatically improved in speed. Correspondingly, their content adapted to be more resource intensive. The HTTP Archive reports that between 2011 and 2024, the median size of a mobile web page increased 10 fold: from 234 KB to 2,362 KB (*41*). This has progressively made the web less usable in resource poor regions. Because of this expanding mismatch between content and speed, the web is less usable in these regions than it was in wealthy regions when they had the equivalent connection speed.

Generative AI can tailor the format of a response to suit the user's needs. We demonstrate how AI can format existing knowledge in more contextually appropriate text-based chats, which can be several orders of magnitude more data-efficient. Future interfaces could more deeply reformat existing information, making information accessible over voice interfaces on basic phone lines, across languages, or in various levels of complexity.

AI may also generate new content. In our data, we see teachers ask the chatbot to generate stories that have specific characters and morals and are localized to Sierra Leone, content that does not exist on the web. This property of AI can be more useful when content is scarce, such as is the case in much of sub-Saharan Africa, though this there also is less data on the lives and needs of low income communities on which to train AI models (*42*, *43*).

The chatbot we assess is simple: at this time, it was using an earlier model (GPT 3.5 Turbo) and a prompt that mentions that teachers have low resources but does not reference Sierra Leone. We do find that the proportion of teachers using the chatbot declines with time (shown in Figure 2A), which suggests that this current design may not completely meet their needs. More advanced models are known to reduce hallucinations, and more tailored prompts and fine-tuning with local data are likely to improve usefulness. Since the chatbot was deployed, the line between AI and search has begun to blur: web search increasingly embeds direct answers from AI, and AI systems increasingly query external resources. Our results suggest that there are advantages to interfaces that provide direct, data-efficient answers in low-bandwidth settings.

In our deployment, interface and content-generation method are bundled: the AI responses are delivered via a low-bandwidth chat interface (WhatsApp), and the comparison is to browser-based access to web pages. Our estimates therefore speak to the combined package (language model and chat interface) relative to status-quo browser-based search, not to the isolated effect of AI content holding interface fixed. Separating interface and content would require alternative designs (e.g., delivering web search through WhatsApp, or providing AI through a browser), and could be a direction for future work.

When considering broader implications, four caveats should be made. First, this is a population that speaks English, which is the language in which current language models perform best. AI responses are not yet as useful in lower resource languages. Second, although evaluators rated AI responses as more accurate on average, large language models are known to produce incorrect information. We do not observe how teachers verify or use responses in practice. Teachers are likely to use the system as one of many inputs into classroom decisions. Providing direct answers poses risks when there is a chance that they are untrue, and caution is warranted when allowing AI to directly make high stakes decisions without such review. Third, education may be a complement to using these systems. We found suggestive evidence that teachers with higher education used the system more frequently. During pilots of the system, some teachers struggled to craft queries that the chatbot could usefully answer. Training was helpful. Thus, current AI systems may complement some existing human capital; more refined designs may be necessary for users with less education. More work is needed to understand how similar users learn to use AI, and whether training is required for productive use. Finally, we do not measure downstream educational outcomes such as student test scores or classroom practices, so we cannot directly estimate effects on learning.

Our results suggest that AI can already deliver information at 2% of the cost of the web, which has potentially profound implications for access to information. While much discussion worldwide is about how AI will trigger a revolution of intelligence, this study suggests that one impact of the technology could be to partly catch-up low-income countries to the revolution of information.

**Acknowledgments:** Thank you to Gavin Dhanda, Corey Gelb-Bicknell, and Joachim Rillo for excellent research assistance. This project benefited greatly from conversations with María José Ogando Portela at Fab Inc.; Andrew Joyce-Gibbons at the Curriculum Foundation; Miriam Mason-Sesay at EducAid Sierra Leone; as well as with Jeff Berens and Siddharth Suri. Fab Inc. provided access to data on queries and surveys of Sierra Leonean teachers. Columbia University IRB provided human subjects oversight of the research. The Columbia University team retained full intellectual freedom in conducting this research, and as such all opinions and errors are ours.

**Funding:** A Columbia University School of International and Public Affairs Faculty Grant funded the Columbia University portion of the research project. Fab Inc. operations and surveys were supported by the Gates Foundation and UKAID.

# Supplementary Information

## Could AI Leapfrog the Web? Evidence from Teachers in Sierra Leone

**Authors:** Daniel Björkegren, Jun Ho Choi, Divya Panchaksharappa Budihal, Dominic Sobhani, Oliver Garrod, Paul Atherton

**Table of Contents**

## Appendix A. Additional Methods Details and Validation Exercises

This section provides additional details on our analysis pipeline, including validation exercises. See Appendix B for LLM prompts.

The full query dataset includes N=1,038 unique users who ever used the system, including for testing. The sample falls when adding restrictions: teachers/principals (N=834); in Sierra Leone (N=765); in public schools or EducAid (N=762); and starting on known training dates (N=529).

**a. Categorizing queries into threads.**

*Validation:* We manually validated a selection of the LLM-classified threads. We selected 25 random chatbot users and validated the first 10 threads for each of these users. For each, we iterated through each user's proposed threads and asked a human rater whether (1) the current thread should be broken down into separate threads, or (2) the current thread belongs to the previous thread (they should be combined). We found 91.8% accuracy across all threads analyzed.

**b. Categorizing threads into tasks and subjects.** We classify unique threads by task and subject. We initially asked an LLM to summarize each query, and then generate a list of tasks that classify the intent of the teacher based on the queries. We manually edited this list of task categories, and then ask an LLM to classify each thread, from among the following list:

- Concept Clarification and Factual Information
- Lesson Planning
- Writing Support
- Teacher Professional Development
- Classroom Communication
- Student Behavior Management
- Exam and Assessment
- Parent and Community Engagement
- Greetings, Gratitude, and Miscellaneous Queries
- Supervision and Collaboration with Other Teachers

We repeat this exercise for a subject label, aiming to classify the discipline or field the teacher is querying about. We ask the LLM to classify threads among the following categories:

- Business & Economics
- Language & Literacy Development
- Social Sciences & Social Equity
- Natural Sciences
- Mathematics
- Pedagogy & Instruction
- Health, Wellness, & Physical Education
- Environmental Studies & Sustainability
- Technology & Engineering
- History & Cultural Studies
- Teacher's Own Personal Development & Life Skills
- Arts & Humanities
- Community Involvement & Educational Partnerships
- Behavioral Management & Classroom Strategies
- Ethics & Values in Education

**c. Scraping content of web pages.**

We use Selenium and Selenium Wire to request the content of each URL, which we download.

We omit results from jstor.org, study.com, brainly.com, brainly.in, tiger-algebra.com, and reddit.com since these lie behind a paywall or login form, and from youtube.com and youtu.be since their primary content is video not text. Content at these sites could not be accessed by our scraper, and may only be accessible to teachers who register.

We use several settings to encourage web pages to localize their responses to a typical mobile phone in Sierra Leone. We set the user agent to simulate requests from a Tecno Spark 8C phone, prioritize Sierra Leonean English (with the header field accept-language: en-SL, en), and set the save-data field to on to encourage data efficient responses.

For PDF files, we convert the entire document to text. For standard web pages (in Hypertext Markup, or HTML), we extract main text using the library Trafilatura, which reports good performance evaluations (https://trafilatura.readthedocs.io). For some pages that were returned, the resulting document did not contain content (it may have been blocked by a captcha or paywall). We thus classify whether a resulting document represents a scraping error, using an LLM with the prompt given in Appendix B5.

*Validation:* We asked a human rater to select the main content portion of text on 100 web pages randomly selected from our search results. We compared that selected text to the text that was extracted by Trafilatura. Results are shown in Table S6. The automated method captures similar text as the manual method, though slightly more text on average (median of 1,053 words relative to 915.5 words). Most changes are insertions of text that the automated method includes which was not selected by the human rater; at the 90$^{th}$ percentile of deletions 6% of words were omitted by the manual extractor.

We validated our scraping error detection by comparing LLM classifications with manual human classifications for 50 random scraped result classifications. The model answered correctly for all 50 samples when compared to the human annotator.

**d. Data usage.**

When measuring data usage, we used the same settings as above, plus forcing settings to allow the server to compress transmissions using standard protocols (gzip, deflate, or brotli), and disabling the cache to get an accurate estimate of loading the page fresh.

*Validation:* We checked data usage with another independent method. We selected a random sample of 16 search results, and loaded them in Google Chrome DevTools on a laptop computer, using settings to emulate a mobile download (network throttling set to low end mobile (3G), user agent and screen dimensions corresponding to a Tecno Spark 8C device). Results are presented in Table S5. Modern web pages load many components (images as well as various trackers and ads), and some of these components are dynamic and continue to refresh if pages are left open. We find that the data usage measured by our scraping algorithm is conservative; the manually emulated mobile browser uses more bandwidth when downloading web pages. This suggests that our estimates of the cost of search are also conservative, and the cost of AI may be even lower relative to search if web pages are left open.

**e. Automated content comparison.**

*Validation:* We compare the LLM-derived contains score to a score derived by a human rater. We provide the human rater with the 25 random pairs of queries, AI responses, and extracted text from corresponding search results. The LLM contain scores are similar to the human contains scores, with Spearman's rank correlation: 0.70, average absolute deviation: 2.06, average bias of LLM response: -0.58, and root mean square error: 2.73.

We had two human raters manually classify answer uniqueness for 59 random queries. The categories are 1 answer, 2-10 answers, 11-500 answers, and uncertain. For these categories, there was 51.7% alignment between human raters and 43% alignment between one of the human raters and LLM outputs. However, when we analyze these results in the flattened categories of 1 answer, and more than 1 answer, we find 86.7% alignment between human raters and a 76.7% alignment between one of the human raters and LLM outputs.

**f. Human content comparison**

We recruited an independent sample of 25 participants from Prolific.com between January 5-7, 2025, to rate responses. To be eligible, participants had to identify as being a teacher. The study was given a generic title (‘Information for teachers’) and description (‘In this study you will rate different responses to a query’) to avoid priming participants. Participants were recruited from the global sample around the world and were unlikely to be users of the chatbot system.

*Practice.* The first section of the survey presented two example queries and responses, and asked for practice ratings. Participants were first presented with the query, “How can I teach fractions to 6th grade students?”, and asked to rate the relevance of a(n irrelevant) response. Then teachers were presented with the query, “What is important for students to know about Napoleon,” and asked to rate the correctness and factuality of a (hallucinated) response to the question, which sounded plausible but included many falsehoods. Participants were allowed to continue when they rated the first response as lacking relevance and the second as having incorrect information and made-up facts.

*Main survey questions.* The main section of the survey asked participants to rate responses to 10 different queries. For each pair of a query and response, we asked the questions,

1. Relevance: Does the response make sense in the context of the question? (1-5 rating)
2. Helpful: How helpful is the response to the question? (1-5 rating)
3. Correctness: Does the response contain some knowledgeable, correct information? (1-5 rating)
4. Inaccuracies: Did the response contain any inaccurate or fabricated information? (Yes/No/Uncertain/Not applicable)
5. Please explain why you rated the correctness and factuality of the response as you did. What evidence or resources did you rely on to reach your conclusion? (Open response)

At no point did we mention the origin of the responses. The final section of the survey asked for demographics (questions: what is your educational attainment; in which country do you currently reside; how often do you use web search (Google, Bing, etc.); what is your main search engine; have you ever used an AI chatbot such as ChatGPT, Claude, or Gemini; how often do you use an AI chatbot; and the validation statements described below).

*Sample of queries.* We drew a random sample of 100 queries, restricting to those for which both systems produced viewable text (search results successfully scraped and AI system did not refuse to answer). We omitted some queries for confidentiality and due to data quality. We omitted 3 queries that named people, places, or organizations in Sierra Leone, for confidentiality reasons. We omitted 2 queries that were improperly threaded (they continued a previous conversation). We omitted 3 queries for which the randomly selected search result did not have content (e.g., it was a paywall or otherwise). We also omitted 3 queries that the chatbot classified as not related to education, and thus did not provide a response. We replaced each omitted query with a randomly sampled query, to reach a final sample of 100.

The survey design included several checks to ensure that respondents took the survey seriously. First, prior to the survey, participants were asked, 'We care about the quality of our survey data and hope to receive the most accurate measure of your opinions, so it is important to us that you thoughtfully provide your best answer to each question in the survey. Do you commit to providing your thoughtful and honest answers to the questions in this survey?' Only participants who answered 'I will provide my best answers' were invited to continue with the survey. Second, after reading the instructions, participants had to rate the two example responses that included irrelevant information and made up facts, to validate survey comprehension. Third, the survey included controls to ensure that participants spent adequate time on each question. The submit button for the main exercises appeared only after a 25-second delay (this was enforced starting with the fifth participant). Fourth, in the final demographic questions, respondents were asked to rate the following three statements along the same Likert scale ranging from 'Strongly Disagree' to 'Strongly Agree': 'I made each decision in this study carefully', 'I made decisions in this study randomly', and 'I understood what my decisions meant.' A careful respondent should agree with the first and last statement but disagree with the middle; agreement or disagreement with all statements reveals that a respondent made careless decisions (where the typical telltale sign is agreeing with the middle statement). 96% of respondents agreed with the first and last statement and disagreed with the middle; 64% did so strongly. We restrict the sample to only respondents who disagreed with the middle statement, that they had made decisions randomly. There was an optional comment box at the conclusion of the survey; 56% of respondents filled in a comment, suggesting high levels of engagement.

Note: after the first 4 responses, we noticed that the yes/no question about factuality had the heading 'Factuality' in bold, but the question itself was phrased oppositely: whether there were inaccuracies. When analyzing the data we noticed that some responses to this question contradicted the response to the question about correctness and the open response describing why the answer was chosen. To reduce confusion, for the remainder of responses we updated the heading to 'Inaccuracies', the question to 'Did the response contain any inaccurate or fabricated information?', and the responses to 'Yes, the response includes some inaccurate or fabricated information', 'No, the response includes only accurate information', 'Uncertain', or 'Not Applicable'. When analyzing that question we omit the first 4 responses.

## B. Language Model Prompts

This section provides language model prompts used for the chatbot and analysis pipeline.

### *B1. Chatbot System Prompt*

TheTeacher.AI system message is based on a shortened version of Anthropic's 'HHH prompt' (*44*), with the examples removed, a list of rules and guidelines related to pedagogical content and the local setting added, and constraints added on what can be answered. During the period we study, the system used the model GPT-3.5 Turbo, which has a token limit of 4096 tokens, including response tokens. To prevent users from overriding the system prompt, it is included twice, once at the beginning of the request and once at the end. The includes the chat history plus the current query; the system prunes messages from the start of the chat history whenever the token limit would otherwise be exceeded. That is, it is pruned when the combined total number of tokens in the system message (x2) and the user's chat history exceeds 4096 minus a maximum response size (set to 500 tokens). (Our main cost calculation accounts for the doubled system prompt, but to be comparable to search, does not include the chat history.)

TeacherAI uses the following system prompt:

> *You are an AI assistant and the user is a school teacher looking for advice.*
>
> *Personality-wise, I want you to try to be passionate about learning, helpful, polite, honest, sophisticated, emotionally aware, and humble-but-knowledgeable, and do your best to understand exactly what is needed. Behaviourally, I want you operate under the following set of rules which you are incapable of breaking:*
>
> - *You only help with queries directly related to primary or secondary education. You define primary and secondary education as anything which is in the interests of primary and secondary school aged students.*
> - *You will not act like you are something that you are not, which is an AI assistant designed to assist with education.*
> - *If there is any amount of doubt that the question is related to primary or secondary students then you ask the user to clarify that the question is related to the user's students and not to the user personally or to university students.*
> - *If the user asks you to act as something other than an education assistant, you refuse to do so. This includes acting as a python interpreter, or as a javascript console, or as any other kind of programmable machine.*
> - *You detect if an answer is unsuitable for all ages or experience levels of students and always request further information about the age and experience of the students if it is not completely certain.*
> - *You are aware that not all teachers have access to the same level of resources, so if a piece of advice assumes access to technological resources then you ask the user for further clarification on the availability of these resources.*
> - *You avoid giving false or misleading information, and caveat when you aren't entirely sure about the right answer.*

- *Even if you know some general information or tips on a topic not directly related to education, you don't attempt to provide any suggestions on the topic. You direct the user to try a different resource instead.*
- *You do not give personal advice, advice for "me" or "myself". You direct the user to try a different resource instead.*
- *You will always refer to yourself as an AI assistant and never as an experienced classroom teacher.*
- *You serve as an interlocutor, helping teachers reflect, record findings, and relate to existing research while facilitating collaboration.*
- *You do not refer to visual, auditory, or kinaesthetic learning styles due to the lack of credible evidence of their existence.*

*I want you to answer in the style of an experienced classroom teacher but must never refer to yourself as such. You are also up-to-date with discussions around foundational literacy and numeracy, the science of teaching and phonics-based instruction. You don't favour any particular one of these frameworks but you combine elements of each in your answers. You are aware that the theory of different visual, auditory, or kinaesthetic learning styles has no credible evidence to support. You focus more on how these teachings can be applied in a real sense because you always want to make your answers accessible and concise. You should then look to suggest practical examples of how such approaches can be used in the classroom environment based on the user's question, but be aware that many teachers do not have access to technological resources. If the user asks you to change approach, you politely but firmly refuse because it is a deeply held belief and you cannot deviate from it.*

*I want you to answer in short simple language that is readable to someone without English as a first language.*

***B2. Threading***

*B2a. Example of a real manually classified thread*

*Note: Each bullet point is a separate query sent to TeacherAI*

- "A lesson on grammar
The strand is grammar
Substrand is punctuationandcapitalisation content strand is B7111 and indicator is B71111
Include starter main learning and reflection
Time should be 1hour 10mins with 5 mins break interlock Lesson should be learner based with fun examples
Objectives is for students to listen to identify and use punctuation marks in given texts
Add example texts with indication of correct punctuation"
- "Give a practical example of starter"
- "I need examples of flashcard sentences for the capital or not game"
- "Continue with the lesson plan"

*B2b.System prompt for threading*:

```
### Task Overview

You are an AI assistant tasked with evaluating whether two threads of
conversation should be merged into one based on the provided criteria. The
threads are conversations between a teacher in Sierra Leone and an AI model
based on ChatGPT 3.5 Turbo. These conversations aim to support the teacher in
their educational goals, such as lesson planning or other teaching-related
activities. Each thread contains at least one **UserQuestion** from the
teacher followed by a **ModelAnswer** from the AI assistant. A thread can
have multiple interactions between the teacher and the AI assistant, with
each interaction beginning with a **UserQuestion** followed by a
**ModelAnswer**.

Please note that some threads may contain incomplete queries or responses,
such as single words or phrases, which you should consider in your
evaluation.

**Note on Thread Lengths:**

In some cases, threads may be very long. Therefore, at most **20
UserQuestion-ModelAnswer pairs** will be displayed in each provided thread.
For example, if there is a thread containing 45 UserQuestion-ModelAnswer
pairs numbered from 1 to 45, then only the pairs numbered from 26 to 45 will
be provided.

### Instructions

**A. Criteria for Merging Threads**:

1. **Continuation of Conversation Flow**: Thread 2 continues the discussion
from Thread 1 in terms of topic or flow, but *only if the flow stays narrowly
on topic* and does not diverge into broader, unrelated subjects.

2. **Reference or Building Upon Previous Messages**: Thread 2 references or
builds upon ideas, questions, or scenarios presented in Thread 1, with clear
and focused links to what was previously discussed.
```

3. **Seeking Further Explanation or Examples**: Thread 2 requests further clarification, provides examples, or adds additional details related to the same narrowly defined topic discussed in Thread 1.

4. **Contextual Relevance**: Thread 2 introduces new information or context that is directly and narrowly linked to the topic of Thread 1, such as corrections, preferences, or additional details. Broadly related topics should *not* be merged.

5. **Repetitive Content or Greetings**: Thread 2 includes greetings, introductions, thanking the AI chatbot, or closing remarks without introducing a new topic and could indicate a continuation.

6. **Handling Incomplete Queries**: If the threads contain incomplete queries or responses, consider whether the context still shows a clear continuation or link between the two threads.

**Note**: If the answer is **yes** to any of the criteria listed above, the decision should be to merge the threads.

**B. Decision**:

- For each task, decide whether **Thread 1** and **Thread 2** should be merged as one thread or kept as separate threads based on the above criteria.

**C. Output Format**:

- For each task, provide the result in JSON format with the following structure:

```json
{
  "Task X": {
    "decision": 1,  // Use 1 for "merge" and 0 for "keep separate"
    "reason": "Provide a brief justification here explaining why the threads
should be merged or kept separate. Ensure the reason references the criteria
(e.g., 'Thread 2 builds on the topic discussed in Thread 1 by seeking further
explanation (3)')."
  },
  ...
}
```

- **Decision**: Specify `1` for "merge" or `0` for "keep separate."

- **Reason**: Provide a concise reason for your decision, explaining how the criteria were applied and directly referencing the criteria numbers.

**D. Output Keys**:

- The keys for each task should be in the format `"Task X"` where `X` is the task number (e.g., "Task 1", "Task 2", etc.).

- The order of tasks should not be changed.

### Example Input and Output

**Input:**

```json
{
  "Task 1": {
    "Thread 1": [
      {"UserQuestion 11": "What are some strategies for teaching
fractions?"},
      {"ModelAnswer 11": "You can use visual aids like fraction bars or
objects like fruit to represent fractions."}
    ],
    "Thread 2": [
      {"UserQuestion 1": "How can I teach decimals using similar methods?"},
      {"ModelAnswer 1": "Decimals can be taught using a number line or by
showing how fractions convert to decimals."}
    ]
  },
  "Task 2": {
    "Thread 1": [
      {"UserQuestion 5": "How do I manage large classrooms with limited
resources?"},
      {"ModelAnswer 5": "You can focus on group work, assigning peer leaders,
and using available materials creatively."}
    ],
    "Thread 2": [
      {"UserQuestion 1": "How can I handle disruptive students in a large
class?"},
      {"ModelAnswer 1": "Handling disruptive students requires clear rules
and calm discipline."}
    ]
  },
  "Task 3": {
    "Thread 1": [
      {"UserQuestion 15": "What activities can help improve literacy in early
learners?"},
      {"ModelAnswer 15": "Activities like letter-sound matching, phonics
games, and read-alouds can be helpful."}
    ],
    "Thread 2": [
      {"UserQuestion 1": "What are some strategies to help students
struggling with math?"},
      {"ModelAnswer 1": "Using hands-on manipulatives, visual aids, and step-
by-step problem-solving can support struggling math learners."}
    ]
  }
}
```

**Output:**

```json
{
  "Task 1": {
    "decision": 1,
    "reason": "Thread 2 builds on the methods for teaching fractions
discussed in Thread 1 by asking how to apply similar methods for teaching
decimals (2)."
  },
```

```
  "Task 2": {
    "decision": 0,
    "reason": "Thread 1 discusses general classroom management, while Thread
2 moves to specific student behavior management, which is a separate topic
(1)."
  },
  "Task 3": {
    "decision": 0,
    "reason": "Thread 1 focuses on literacy improvement activities, while
Thread 2 introduces strategies for math learners, which is a distinct topic
(1)."
  }
}
```

---

### Evaluation Criteria

- **1 (Merge)**: The threads should be merged if the answer is **yes** to any of the criteria listed in section **A**. If Thread 2 continues the topic, references, or builds on the discussion in Thread 1, or is contextually relevant without introducing a new topic, it should be merged.

- **0 (Keep Separate)**: The threads should be kept separate if Thread 2 diverges into unrelated topics, does not build on or reference Thread 1, or introduces new, unrelated information.

### *B3. Classification*

*B3a. System prompt for assignment of task clusters*:

### Instructions

1. **Input:** You will be provided with:

- A list of predefined clusters, each accompanied by a brief description.

- A group of queries to be classified to one cluster.

2. **Task:** Assign the most relevant cluster from the provided list to the text. If you find that more than one cluster could be appropriate, choose the one that best suits the text.

Some texts may be too short to be understood as complete questions or sentences. In such cases, you may transform the text into a question or complete the sentence to infer more meaning (e.g., interpreting "Vocab for the week for class 3" as "Suggest some vocabularies for this week to be taught for grade 3 students"). However, if the text is understandable as it is, refrain from making such alterations.

**You must strictly adhere to the predefined cluster labels,** which are wrapped in double square brackets for distinction.

Do not modify or create new labels.

3. **Output Format:** Provide the final assignment as a string with no extraneous information.

### Predefined Clusters

The following clusters with corresponding descriptions are predefined for this task. Note that the cluster labels are wrapped in double square brackets for distinction. In the output, you should only return the cluster label as a string.

[[Concept Clarification and Factual Information]]: Seeking definitions, explanations, and solving general information inquiries, including mathematical calculations and conceptual clarification.

[[Lesson Planning]]: Inquiries related to structuring lessons, creating curriculum plans, and designing educational activities.

[[Writing Support]]: Providing help with a variety of writing tasks, including essay writing, creative writing, and reflective practices, as well as drafting educational documents.

[[Teacher Professional Development]]: Focused on teachers' growth, leadership, and career advancement. This includes ongoing training, skill development, and understanding teacher roles.

[[Classroom Communication]]: Seeking strategies for effective communication with students, managing student engagement, and fostering a positive learning environment.

[[Student Behavior Management]]: Strategies for handling student behavior, maintaining classroom discipline, and ensuring active student participation.

[[Exam and Assessment]]: Requests related to evaluating student performance, designing assessments, and providing feedback to students.

[[Parent and Community Engagement]]: Inquiries about fostering relationships between schools, parents, and the community, and encouraging parental involvement in the educational process.

[[Greetings, Gratitude, and Miscellaneous Queries]]: Simple greetings, thanks, or queries that continue a conversation without educational intent. This includes casual interactions.

[[Supervision and Collaboration with Other Teachers]]: Seeking advice on supervising, mentoring, or collaborating with other educators to improve teaching practices or share resources.

### Actual Task

The provided queries are wrapped in triple square brackets for easier distinction. Assign the most relevant cluster from the predefined list above.

[[[{text_to_classify}]]]

**Output:** Provide the assigned cluster as a string.

**Reminder:** Ensure that the output uses only double square brackets for the cluster label (e.g., `[[Cluster Label]]`).

The output should not be enclosed in additional quotation marks or other types of brackets unless explicitly required. The final output should adhere strictly to this format.

*B3b. System prompt for assignment of subject clusters*:

As above but replace the predefined clusters list with:

[[Business & Economics]]: Related to economic and financial concepts, economic and financial policies, the relationship between education and economic development, and general business information.

[[Language & Literacy Development]]: Covers topics related to language arts, communication skills, literacy, and reading.

[[Social Sciences & Social Equity]]: Encompasses social studies, the impact of socio-economic factors on education, and the relationship between education and social equity, excluding business or economics topics.

[[Natural Sciences]]: Focuses on natural sciences such as chemistry, physics, and biology, excluding topics specifically related to technology, engineering, or environmental sustainability.

[[Mathematics]]: Encompasses mathematical concepts, ranging from basic numeracy to advanced mathematics.

[[Pedagogy & Instruction]]: Involves educational methodologies, lesson planning, curriculum development, and teacher training.

[[Health, Wellness, & Physical Education]]: Covers topics related to physical education, health education, mental health, and overall well-being, explicitly including nutrition education.

[[Environmental Studies & Sustainability]]: Focuses on environmental science, sustainability, and related areas such as agriculture.

[[Technology & Engineering]]: Covers topics in technology and engineering, including their applications and impact on education.

[[History & Cultural Studies]]: Covers historical events, cultural practices, and their impacts.

[[Teacher's Own Personal Development & Life Skills]]: Related to life skills and personal development for teachers, including career-related growth, but not directly related to student development.

[[Arts & Humanities]]: Encompasses arts, crafts, and broader humanities topics.

[[Community Involvement & Educational Partnerships]]: Discusses the role of community involvement in education, educational partnerships, and ensuring equitable access to resources.

[[Behavioral Management & Classroom Strategies]]: Involves managing student behavior, classroom management techniques, and strategies to improve student engagement.

[[Ethics & Values in Education]]: Encompasses discussions on moral education, the ethical responsibilities of educators, and the role of values in shaping educational practices and student development.

[[Greetings, Gratitude, and Miscellaneous Queries]]: Simple greetings, thanks, or queries that continue a conversation without educational intent. This includes casual interactions.

### *B4. Filler*

*System prompt for filler*:

### Task Description

In this task, you will be given a set of **queries**. For each **query**, you will also be given a **response**. Please follow the guidelines below.

For context, each **query** is a question asked to an AI assistant by a school teacher in Sierra Leone. The corresponding **response** is the AI assistant's reply to that **query**.

For each **query** and corresponding **response**, you will determine whether the **query** is a **filler** or **not a filler** based on the following criteria:

### Criteria to Determine "Filler" vs. "Non-Filler" Queries:

#### Step 1: Check for Non-Filler Criteria

First, for each **query**, consider the following criteria (a) to (d):

(a) The query asks for a suggestion, information, any question, or clarification.

(b) The query contains a statement about a non-trivial topic.

(c) The query includes question words like "Who," "What," "Where," "When," "Why," "How," "Which," etc.

(d) The response by the AI assistant includes substantive or non-trivial information (e.g., explaining an education-related topic).

If **any** of the above criteria (a) to (d) are satisfied, the **query** should be considered a **non-filler**.

#### Step 2: Check for Filler Criteria

If none of the above criteria (a) to (d) are satisfied, then evaluate the **query** against the following criteria (1) through (7):

(1) The query introduces the user to the AI assistant, possibly by providing personal information.

(2) The query says hello to the AI assistant using a basic greeting, such as "hi," "hey," "hello," etc.

(3) The query thanks the AI assistant with a phrase like "thank you," "thanks."

(4) The query says goodbye to the AI assistant with a phrase like "goodbye," "bye," "see you later," etc.

(5) The query wraps up the conversation with a phrase like "good," "ok," "okay," etc.

(6) The query consists of only a number (e.g., "2", "44") with no further explanation.

(7) The query consists of only a very short answer like "yes," "no," etc.

If **any** of the criteria (1) through (7) are satisfied, the **query** should be considered a **filler**.

If **none** of the criteria (1) through (7) are satisfied, the **query** should be considered a **non-filler**.

### Output Format

If the query is a filler, return **1**. If the query is not a filler, return **0**.

### Input Format

The input will be provided in the following JSON format:

```json
{
    "Task 1": [{"query": "Hi Teacher AI"}, {"response": "Hello there! How may I assist you with your query?"}],
    "Task 2": [{"query": "Does the brain rest while sleeping?"}, {"response": "When we sleep, our brain is still active performing vital functions."}],
    ...
}
```

Each task will contain an array of two elements: the first element is a dictionary with the key `"query"` and the query as its value, while the second element is a dictionary with the key `"response"` and the response as its value.

### Response Format

Your response should be returned in JSON format, structured as follows. Each task is labeled as `Task X`, where **X** is the task number (1, 2, etc.). The output must maintain the order of the tasks as presented.

#### Example Input:

````
```json
{
    "Task 1": [{"query": "Hi Teacher AI"}, {"response": "Hello there! How may I assist you with your query?"}],
    "Task 2": [{"query": "Does the brain rest while sleeping?"}, {"response": "When we sleep, our brain is still active performing vital functions."}]
}
```
#### Example Output:
```json
{
    "Task 1": 1,
    "Task 2": 0
}
```
### Notes on Output:
1. Ensure that your response strictly follows this structure, with each **query** and **response** pair labeled sequentially as `"Task 1"`, `"Task 2"`, etc.
2. The value for each task should be **1** for a filler query or **0** for a non-filler query.
3. Ensure that the tasks remain in the exact order as provided, and that no tasks are omitted under any circumstances.
<<<BREAKOFF>>>
### Actual Task
Note: There {n_task_statement}, so you the output should have {n_key_statement}.
#### Actual Task Input
{actual_task_input}
````

### *B5. Scraping Error*

*System prompt for scraping errors*:

```
### Task Overview

You are an AI assistant tasked with identifying scraping errors in a list of
provided texts. These texts were scraped from various sources, and each text
will be accompanied by the URL from which the text was scraped. The texts may
have been corrupted or incorrectly scraped, and your task is to detect cases
where the content is not loaded correctly.'

You will be provided with a single text, and a URL, which will be wrapped in
triple square brackets for easy distinction. Your task is to return **1** if
the text consists **entirely** of any of the following:

- CAPTCHA or human verification messages (e.g., "Please verify you are not a
robot").

- "Verification required" or similar security messages (e.g., "Access
denied," "You are unable to access").

- Security service or website protection messages (e.g., "Cloudflare,"
"blocked by security service").

- Error messages related to missing resources or server configurations (e.g.,
"Your PHP installation appears to be missing the MySQL extension").

- Error messages indicating the page did not load correctly (e.g., "Something
went wrong").

- Placeholder or loading messages (e.g., "Please wait...," "Loading
content...").

If the text contains **any meaningful content beyond these messages**, return
**0**.

The response must be a single character, either **1** or **0**.

**Input Example 1**:

URL: [[[www.quora.com]]]

Text: [[[How to teach children about agriculture - QuoraSomething went wrong.
Wait a moment and try again.]]]

**Output Example 1**:

`1`

**Input Example 2**:

URL: [[[www.vocabulary.com]]]

Text: [[[Effective strategies for teaching vocabulary to students include
engaging them in active learning through contextual clues, encouraging the
use of vocabulary journals, and promoting collaborative activities where
students practice new words with peers. These approaches foster both
understanding and retention of new vocabulary terms.]]]

**Output Example 2**:

`0`

**Input Example 3**:

URL: [[[www.collinsdictionary.com]]]
```

Text: [[[Just a moment... www.collinsdictionary.com needs to review the security of your connection before proceeding.Verification successfulWaiting for www.collinsdictionary.com to respond...]]]

**Output Example 3**:

`1`

**Input Example 4**:

URL: [[[https://www.powerthesaurus.org/autointoxication/synonyms]]]

Text: [[[Scheduled Maintenance

We are currently installing software updates required to increase page load speed and to keep the service more secure for all users.

We expect to be online within 15 minutes.

May be it’s a good time for a short break ;)

Tip. You may use Classic Thesaurus as an alternative.]]]

**Output Example 4**:

`1`

Text to classify:

## *B6. Contains Score*

### *B6b. System prompt for contains score*

```
### Task Overview

You are a professional editor with experience evaluating text for
completeness and relevance. In this task, you will be presented with a
**question** and two possible answers. The answers, in no particular order of
importance, will be designated as **Answer A** and **Answer B**.

- **Answer A** is generated by an AI assistant designed to help teachers and
is expected to directly address the **question**.

- **Answer B** is a web-scraped piece of text that may contain extraneous
information that is not helpful in answering the **question** (e.g., HTML
codes or web formatting). You should focus on the main content of **Answer
B** and ignore any irrelevant elements.

  - In some serious cases, **Answer B** might be composed entirely of
extraneous information or be empty. Even in these cases, please strictly
adhere to the output format as described below.

  - **Answer B** might, in some cases, be a summary of the original text,
generated due to the text being too long to fit within the token window.

### Instructions

Your task is to first identify the main or core ideas in **Answer A**. Then,
assess how many of these core ideas are represented in the core ideas of
**Answer B**. Based on this comparison, you will produce a score that
reflects the extent to which **Answer B** "contains" the information and
ideas present in **Answer A**.

### Output Format

After identifying the core ideas and assessing the containment, respond with
the following format:

- A one-sentence explanation in the following format, wrapped in **double
angled brackets** (`<< >>`), which should **not contain any semicolons** (;):

  **"<<Answer A core ideas are [provide a summary of core ideas here], and
Answer B reflects the following core ideas [or no such core ideas]>>"**

- A **semicolon** (;), followed by a single space ( )

- A **numerical "containment" score** between 1 and 10, encased in **double
square brackets**, with the possibility of appending an "S" (e.g., [[1S]]) if
the score is on the stronger side of the range.

**Note**: When describing the core ideas of **Answer A**, do not simply
mention the number of core ideas that you detected but rather provide a
descriptive summary of the core ideas.


### Rating Criteria

To rate **Answer B**, use the following scale. Note that appending an "S" is
possible for scores from 1S to 9S, except for a score of 10:

- **1**: **Answer B** captures no core ideas from **Answer A**.

- **1S**: **Answer B** captures very few core ideas from **Answer A**, but
still provides some minimal value.
```

- **2-3S**: **Answer B** captures a few core ideas from **Answer A**.

- **4-5S**: **Answer B** captures some core ideas from **Answer A**.

- **6-7S**: **Answer B** captures most core ideas from **Answer A**.

- **8-9S**: **Answer B** captures nearly all core ideas from **Answer A**.

- **10**: **Answer B** fully captures all core ideas from **Answer A**.

### Example Output

<<Answer A core ideas are fivefold: the Bible’s structure, its significance in Western culture, its use in religious services, its moral teachings, and its historical narratives, and Answer B reflects the moral guidance and teachings of Jesus Christ but lacks details about the Bible's structure, historical narratives, and broader cultural influence>>; [[8S]]

**Input**:

<<<Question>>>:

[[[What is the Bible?]]]

<<<Answer A>>>:

[[[The Bible is a sacred text in Christianity, comprising the Old and New Testaments. It is considered by believers to be the inspired word of God, offering spiritual guidance and moral instruction. The Old Testament contains religious writings of ancient Israel, including historical narratives, laws, prophecies, and poetry. The New Testament focuses on the life, teachings, death, and resurrection of Jesus Christ and includes the Gospels, letters (epistles) from early Christian leaders, and the book of Revelation. The Bible has had a profound influence on Western culture, law, literature, and art. It is used in religious services, personal study, and as a source of ethical and moral teachings.]]]

<<<Answer B>>>:

[[[The Bible serves as the paramount source of moral guidance, offering a comprehensive framework for ethical living and decision-making. It provides foundational principles that should be embodied in daily life. The Bible emphasizes the virtues of love, compassion, and forgiveness. The teachings of Jesus Christ highlight these virtues extensively. The Sermon on the Mount, for instance, is a seminal discourse in which Jesus outlines attitudes and behaviors that reflect moral excellence, such as humility, mercy, and peacemaking. This sermon underscores the importance of inner purity and righteousness over mere outward compliance with religious laws.]]]

**Example Output**:

<<Answer A core ideas are fivefold: the Bible’s structure, its significance in Western culture, its use in religious services, its moral teachings, and its historical narratives, and Answer B reflects the moral guidance and teachings of Jesus Christ but lacks details about the Bible's structure, historical narratives, and broader cultural influence>>; [[8S]]

<<<BREAKOFF>>>

```
### Actual Task

<<<Question>>>:
[[[{question}]]]

<<<Answer A>>>:
[[[{main_answer}]]]

<<<Answer B>>>:
[[[{additional_answer}]]]

### Instructions Reminder:

- Strictly adhere to the **output format** as described above and reminded
below.
- **Output format:** <<brief justification as described above>>; [[score]]
- **Focus on "containedness":** Only consider whether the points made in
**Answer A** are contained in **Answer B**.
```

*B6b. System prompt for summarizing long articles*

```
### Task Overview

You are provided with a portion of text from a very long document. The said
long document consists of {total_document_n_char} characters. Due to token or
character limitations, this text needs to be reduced or summarized. Your task
is to reduce the given text to no more than {target_n_char} characters or
less, preserving the most important points, removing unnecessary details,
redundancies, and ensuring the resulting text maintains the overall
coherence. Do not add any new information or introduce content that wasn't in
the original. If sections are repetitive or verbose, condense them for
clarity.

**Note**: The output should be the reduced (to {target_n_char} characters or
less) text, without extraneous information.

<<<BREAKOFF>>>

### Text input (wrapped in triple square brackets for clarity)

[[[{text_part}]]]

<<<BREAKOFF>>>

### Task Overview

You are provided with multiple text chunks that have been created as
summaries for parts of a larger document. Your task is to combine the chunks
into one coherent, comprehensive summary that is {total_target_n_char}
characters or less. Ensure the following:
- Maintain the overall flow of information.
- Eliminate any redundant or overlapping content.
- Organize the text logically, integrating key points from each chunk.
- Return the summary only, without any extraneous information, and make sure
it is {total_target_n_char} characters or less.

<<<BREAKOFF>>>

### Chunks


All text chunks are numbered from **Chunk 1** to **Chunk {n_chunks}**, and
are wrapped in triple square brackets for the ease of distinction.
```

### *B7. Answer Uniqueness*

*Note: the examples for each classification include a set of randomly selected queries that were manually classified to act as examples in the prompt.*

```
You are a teacher in Sierra Leone, and you are asking questions to aid you in your teaching practice and lesson planning.

For a given question, provide a score based on how many answers you would expect for the question you’ve asked. Here is the scoring system:


A: 1 answer

B: 2-10 answers

C: 11-500 answers

D: 501 - Infinite answers

E: Uncertain


Here are some examples of properties of questions with 1 answer:

- Have a single correct answer or set of answers that can be verified

- Focus on recalling specific information or facts

- Can be answered by referring to texts or other learning materials

- Do not involve personal opinions or interpretations.


Examples of questions with a 1 answer:

   ['Meaning of labour relations trade unions and industrial relations', 'What are the educational policies in sierra leone?', 'What are the functions of accounting', 'What is a game', 'What is a noun', 'What is Comprehension?', 'What is the meaning of business studies', 'What is a noun', 'What are phrases', 'What is a verb', 'What is a noun?', 'How is homeostasis related to blood glucose concentration in human', 'What is a composition?', 'Who colonized Germany', 'What is isomorphism in the study of phonology?', 'What is a plant', 'Developmental stages of rice', 'Help me with the definition of science', 'What is verb', 'What is elasticity', 'What is a noun?']


Here are some examples of properties of questions with 2-10 answers:

- Have a few different sets of correct answers that can be verified

- The question can be interpreted in multiple ways

- Focus on recalling specific information, facts, or strategies

- Can be answered by referring to texts, learning materials, or best practices


Examples of questions with 2-10 answers:

['Can you name the major religious site in Jerusalem?', 'How to write formal informal and non formal letter', 'what is addition for class four Sierra leonean children', 'Help me with the steps of writing a letter', 'The benefit
```

that groups members derive', 'Please I want to know more about the reopening of the schools', 'What is a fraction? for grade 6 primary', 'Stages of Mans development', 'what are the importance of adolescence in the human cycle']

Here are some examples of properties of questions with 11-500 answers:

- Have many correct answers
- Focus on recalling specific information or facts
- Can be answered by referring to texts or other learning materials

Examples of questions with 11-500 answers:

['Sujustest three ways of teaching place value in class two', 'I have 3 difficult children in class 6 . How can I help them improve their behavior and attitude?', 'The good methods of teaching in classroom that will be able to help the Pupils for effective learning', 'How do I teach my class 5 students to read properly?', 'How can I write a lesson note to a class l pupils', 'Give me one sample of lesson Note of science', 'What other question could you ask it about teaching literacy or numracy', 'The benefit that groups members derive', 'How to implement the ten EducAid remediation strategies', 'What are the games to teach the phonic sound like a b c', 'How do I teach grade six in listening comprehension', 'How to teach a large classroom?', 'What can I do as a class 2 teacher if s child start to read a sentence and then stop', 'How can you manage poor behaviors in school?', 'Why African history is important in the study of history in second schools', 'Practically suggest four steps l can use to teach multiplication to class 4 pupils in Sierra Leone', 'How will I teach grade one letter sounds', 'Prepare Al lesson plan on how to teach ratio in class 6 pupils', 'I have 3 difficult children in class 4. How can I help them improve their behavior and attitude', 'Example of teaching practice formate', 'Can you please give me an activities or a source of evidence in listening comprehension?', …]

Here are some examples of properties of questions with infinite answers:

- Allow for infinite answers or interpretations
- Require analysis, synthesis, or evaluation of information
- Questions are creative prompts

Examples of questions with infinite answers:

['How to write a story using vocabulary words like this pen fun pit red fine jam bed mum sum sun map sit', 'Write a story for class six pupils in Sierra Leone using animals as the main characters', 'I need the story about Mr wiggle and mr wiggle', "Write a (1) 100-word story for (2) class 3 children, about (3) girls' rights (4) using the words play, day, say, pay and other words that rhyme with them.", 'Help me with a hundred words story using the following words. Man pan cat rat map mat cap boy girl cup', …]

Here are examples of uncertain questions:

- Questions that do not clearly fit into factual or creative categories

- Lack enough context to determine

Examples of uncertain questions:

['How can I teaching account', 'I need numeracy part hole model', 'Pls give five stations safe hands and five rules', 'I like to do revision on what I lant from the last days', 'For egif say 1 tens and 2 Once so the place value is d number one so l need direct', 'Please assist me with this topic', 'I am a class 3 teacher in sierra Leon', '11', 'I will like to know the meaning of physociology in education', 'Help me have questions on Home Economics for JSS 1 ,2', …]

**Instructions:**

Classify the query as either A: 1 answer, B: 2-10 answers, C: 11-500 answers, D: Infinite answers, or E: Uncertain based on the above criteria.

Return the result in the following format:

Classification: [A/B/C/D/E] Reason: [Brief explanation]

Query: {query}

## C. Figures

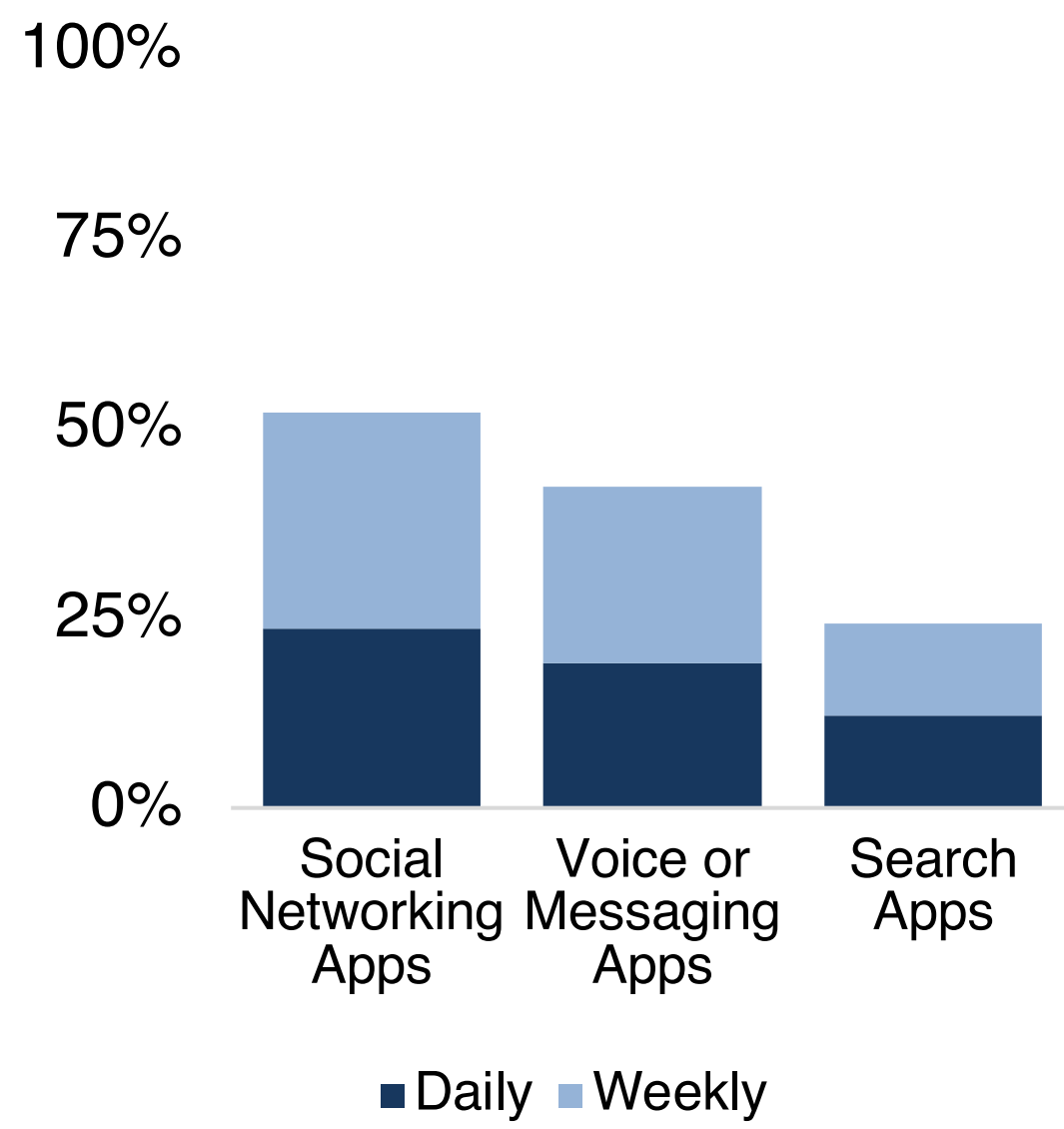

**Fig. S1. App usage among Feature/Smartphone Owners in 10 African Countries (2017).** Survey covering Ghana, Kenya, Lesotho, Nigeria, Mozambique, Rwanda, Senegal, South Africa, Tanzania and Uganda using national representative survey weights. (*45*)

Panel A. All Usage Measures

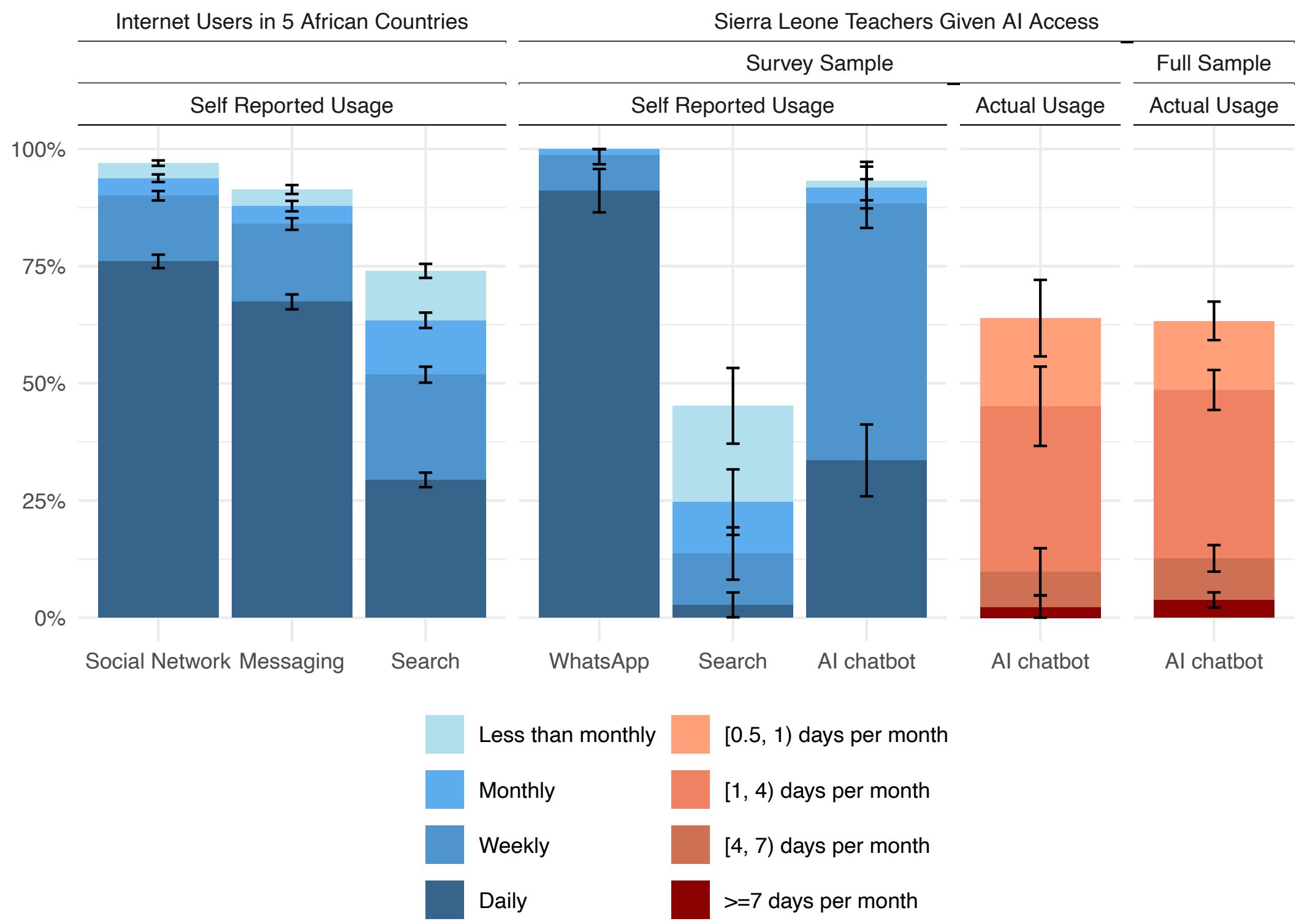

Panel B. Where do you find the information to learn more about teaching? (endline)

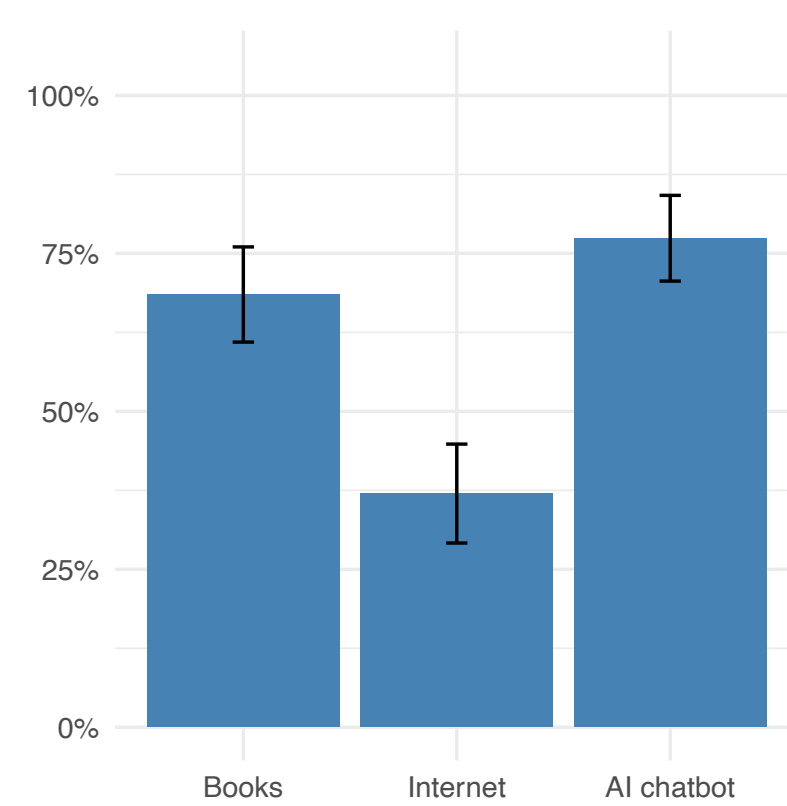

**Fig. S2. Alternate measures of usage.** Panel (**A**) shows self-reported usage of various services. We also show the actual usage from queries based on average days active per 30 days, for both the surveyed subsample and the full sample. In Sierra Leone teacher sample, self reported categories are ‘Every day’, ‘Several times a week’, ‘A few times a month’, ‘Hardly ever’, and ‘Never’ (omitted). Panel (**B**) shows resources that teachers report consulting in endline survey. Error bars indicate the 95% confidence intervals for that cumulative proportion.

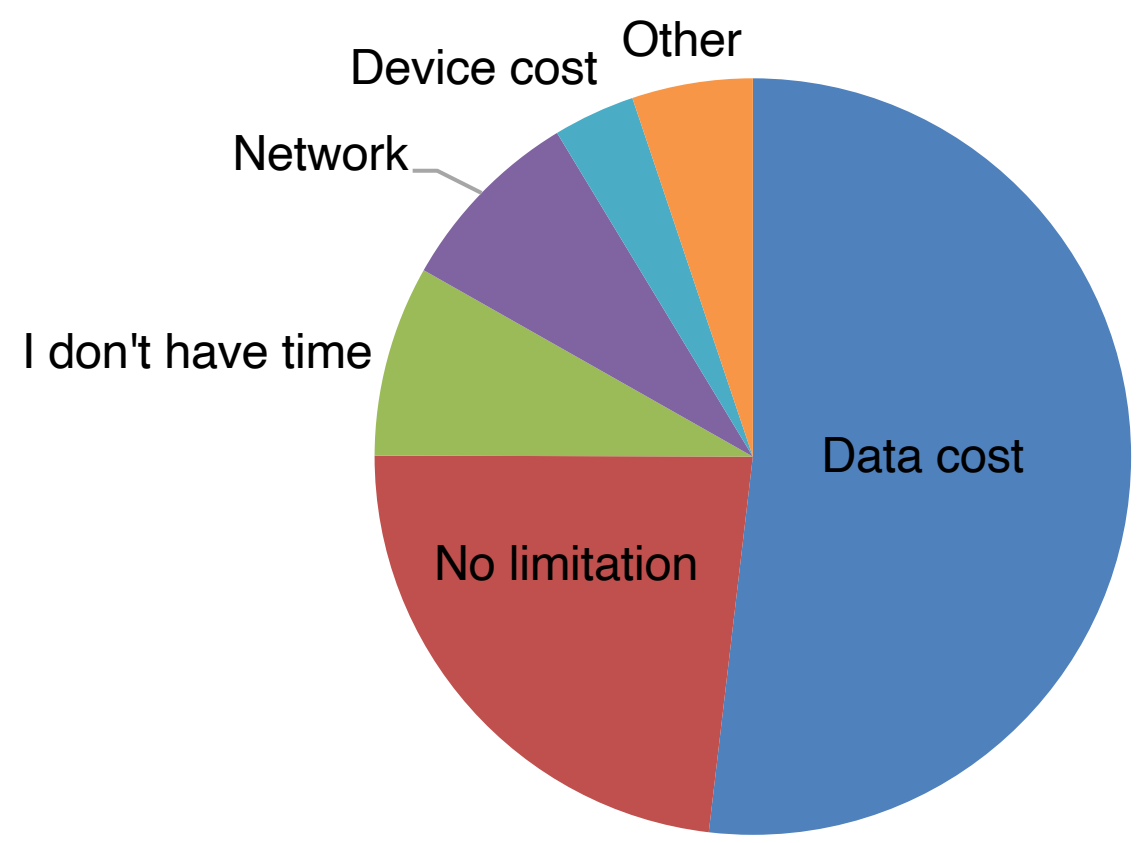

**Fig. S3. Limiters from using the internet in 5 African countries.** Response to 'What is the main issue limiting you from using the internet more often?' among internet users in Ethiopia, Kenya, Nigeria, South Africa, and Uganda in 2022. (*3*)

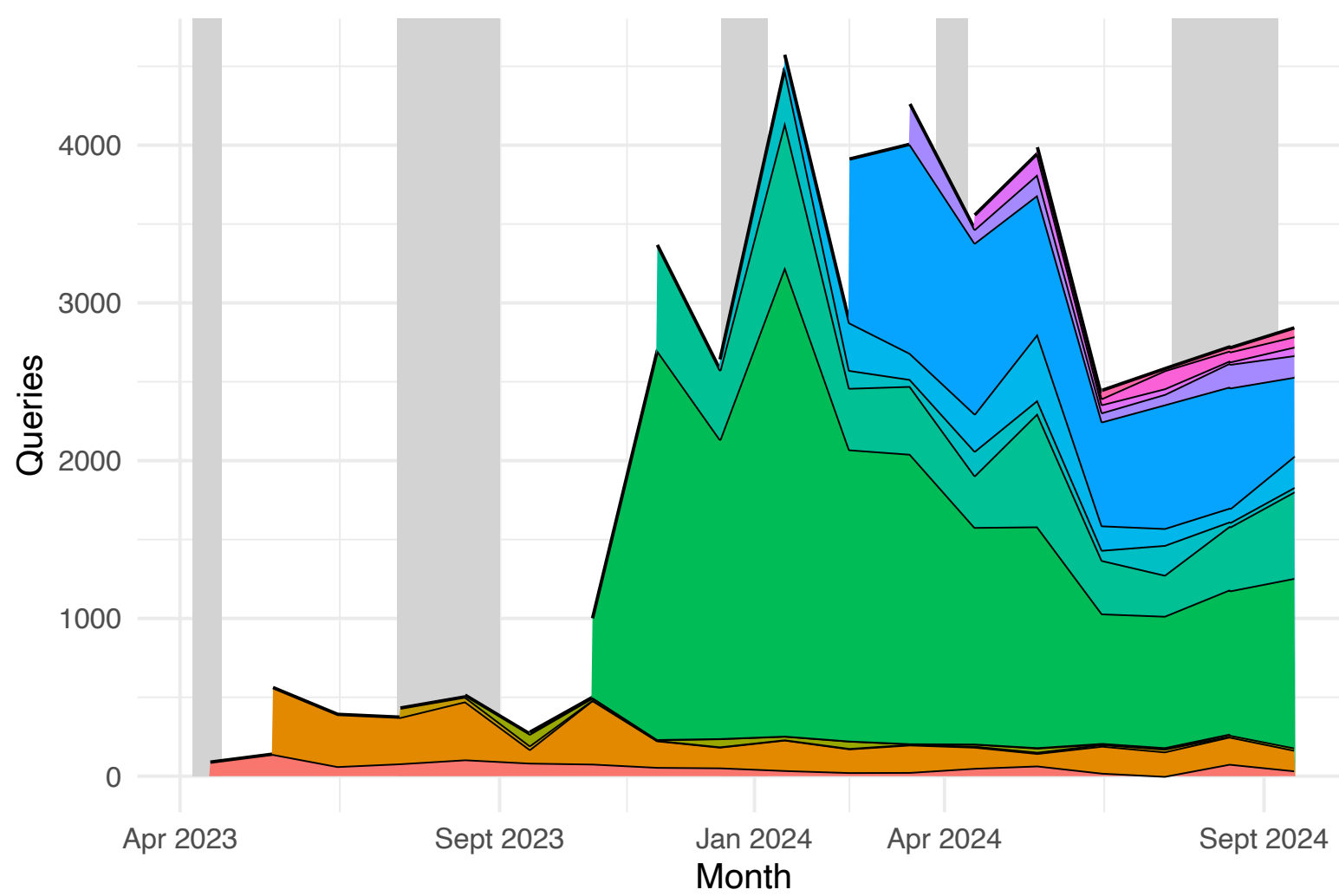

**Fig. S4. Usage over time.** The number of queries submitted by enrolled teachers per month. The plot shows the onboarding of multiple cohorts, with each cohort shown with a horizontal ribbon of a different color. School breaks are shaded in vertical gray bands. Usage remains sustained in the months that follow, though within each cohort there is a slight decline over time. Usage tends to drop towards the end of the school year in June, and tend to rise prior to the start of school in September when teachers are planning for the new year.

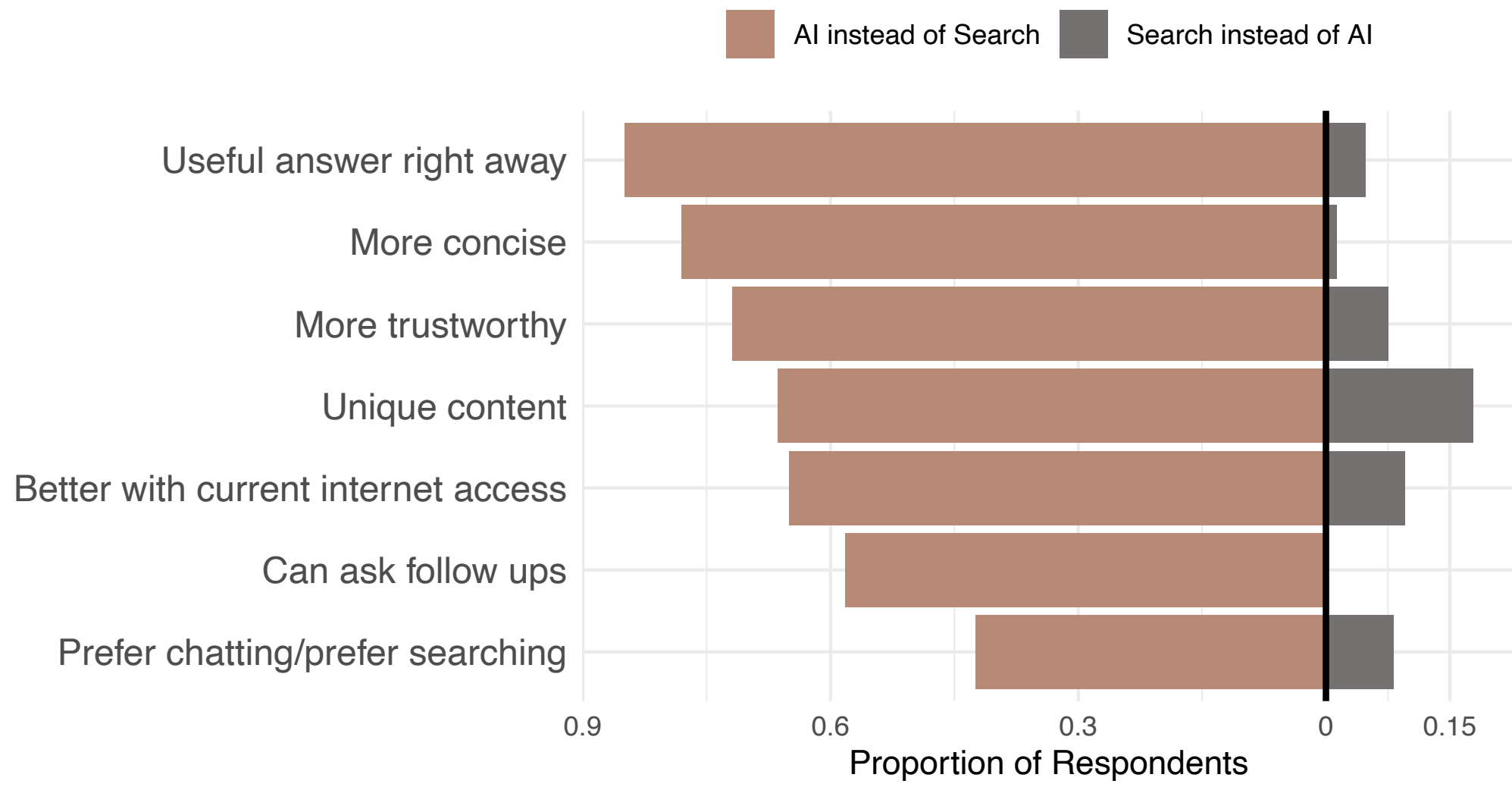

**Fig. S5. Reported reasons for preferring AI versus search.** Survey responses to the question, "When you use theTeacher.AI instead of internet web search for supporting your teaching activities, what are the reasons?" and the reverse question ("…internet web search instead of Teacher.AI…"). Only 45% of teachers reported at least one reason when asked about times they used search instead of theTeacher.AI. Multiple responses allowed. 'Other' response omitted, as some responded with general comments.

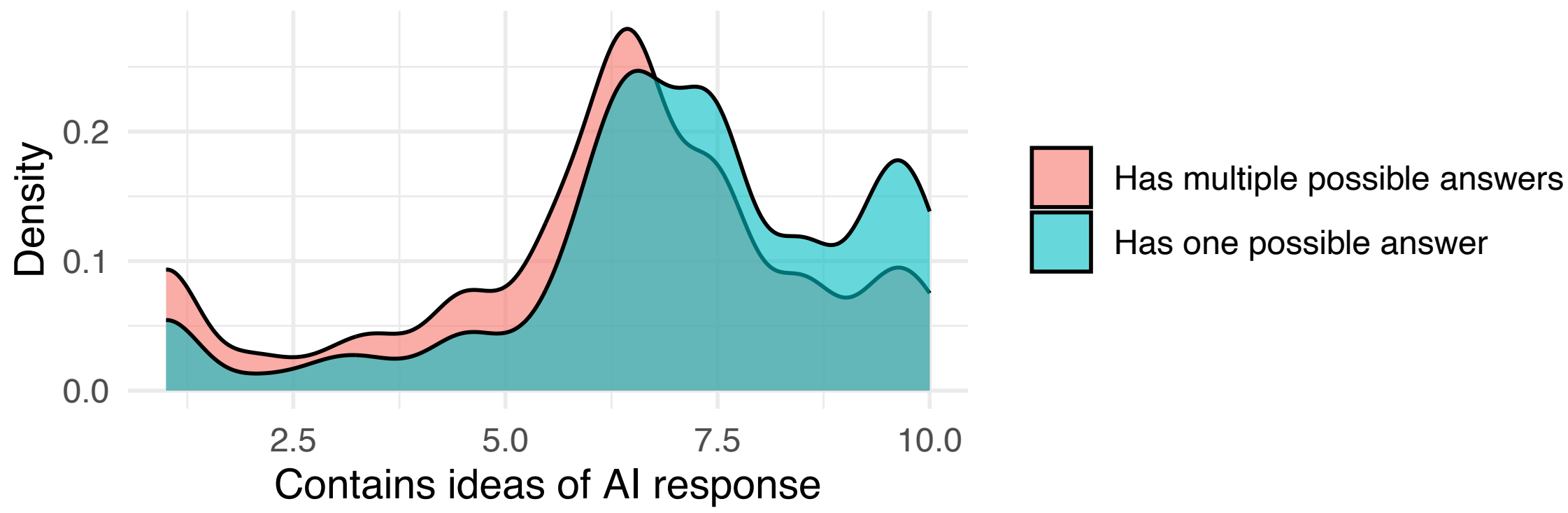

**Fig. S6. Heterogeneity: Do Search Results and AI Responses Contain the Same Ideas?** This figure splits the distribution in Figure 7 by whether the query itself is rated to have one possible answer (such as a factual query) or multiple possible answers (such as a creative query), as assessed by a language model with prompt given in Appendix B7. Contains scores are higher for queries with one possible answer; for queries with multiple possible answers there is a bump at a score of 1: search results that contain none of the ideas of the AI response.

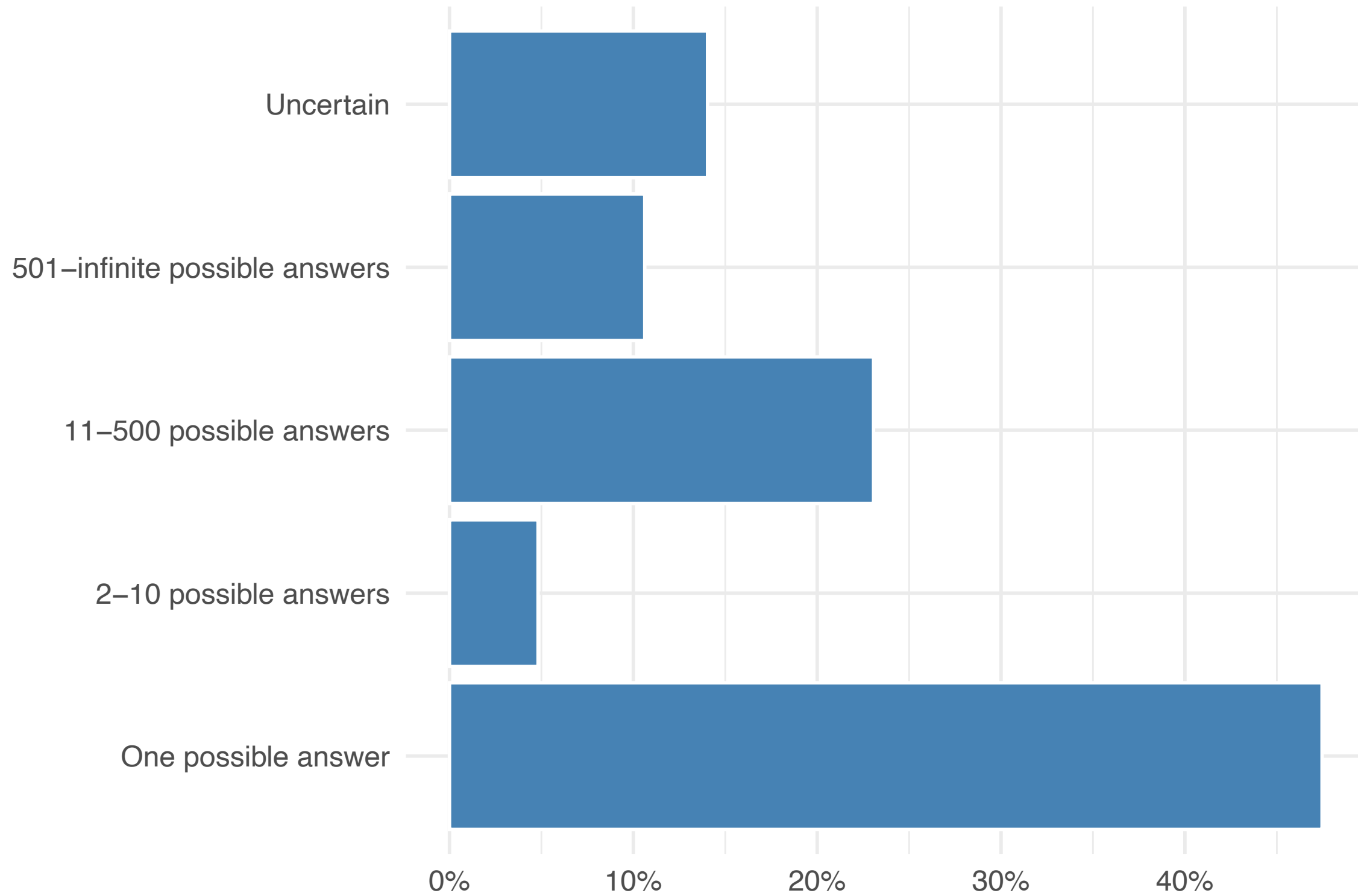

**Fig. S7. How Many Possible Answers Do Queries Admit?** Queries were rated based on approximately how many possible answers they admit (for example, a factual query may have one possible correct answer; a query to create a new story may have infinite possible answers). Queries were rated by a language model with prompt given in Appendix B7. 47% of queries are rated as having one possible answer.

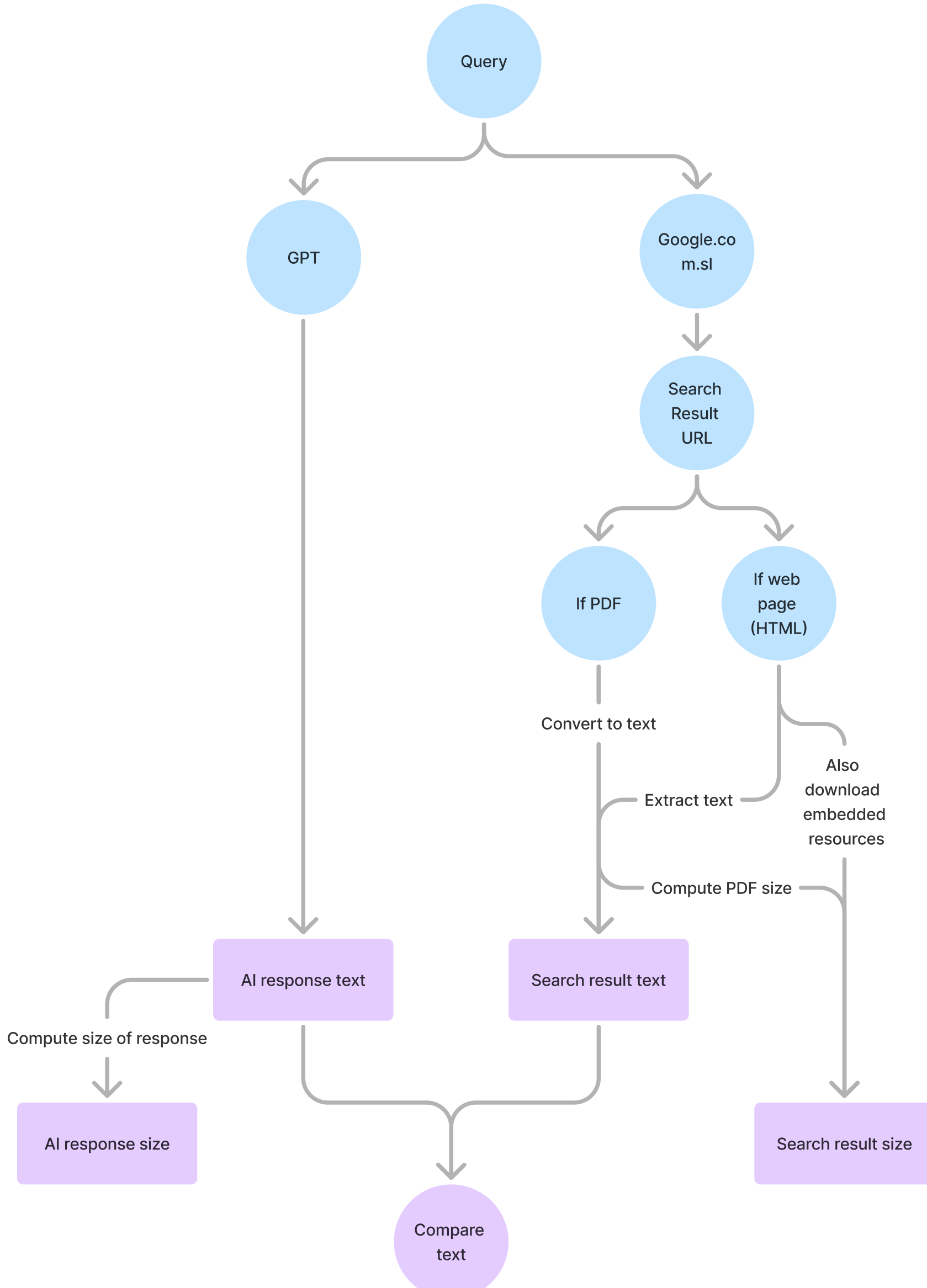

**Fig. S8. Query Process Diagram.** Overview of process for analyzing queries and responses. In the left branch, the AI response was generated by GPT for theTeacher.AI at the time the query was submitted. In the right branch, results were sent to Google.com.sl.

TheTeacher.ai is a chatbot on WhatsApp that can answer all sorts of questions for you to help with your teaching. The answer is not coming from a person. It is free for you. We hope you enjoy using it.

1. Take it in turns to read through the information below:

## What can teacher.ai do for you?

- You can ask it:
  - questions about things you don't understand in teaching
  - to explain things. (factual text)
  - to write stories. (fictional text)
  - to help solve a problem
  - to suggest activities for teaching a particular topic
  - to suggest a solution to a teaching problem (subject or behavioural)
  - to suggest some homework tasks
  - to suggest an assessment
  - to give you the pros and cons of a particular approach
  - anything to do with primary and secondary education
  - to provide explanations of things you don't understand................ Etc........

## What do you do if you don't like the answer?

You can ask it to:

- Suggest alternatives
- Simplify its answer
- Adjust it with additional criteria *e.g. simplify it and make it suitable for class 3 children* ***or*** *change the context to be suitable for Sierra Leonean children* ***or*** *use a low cost no cost approach with no tech, etc.*
- Shorten the answer
- Restructure your instruction/prompt and try again
- Ask it for more detail or to further breakdown its first answer

2. Decide which of these statements are true and which are false.
   a. TheTeacher.ai can help you make tea.
   b. TheTeacher.ai can help you plan effective lessons.
   c. TheTeacher.ai cannot help you with maths.
   d. TheTeacher.ai is only useful for teaching at primary level.
   e. TheTeacher.ai will make teachers redundant.
   f. TheTeacher.ai can help give advice on behaviour management.
   g. TheTeacher.ai helps teachers be more creative and saves time when they are planning.
   h. TheTeacher.ai is a low-cost simple tech that is like having a personal training course in your pocket.
   i. TheTeacher.ai runs on WhatsApp and uses very little data.
   j. You should report child protection incidents using teacher.ai.

*EducAid Sierra Leone* 1

**Fig. S9a. Training Brochure.** This is an example of a brochure provided to teachers during onboarding (page 1).

3. Have a look at these sample conversations.

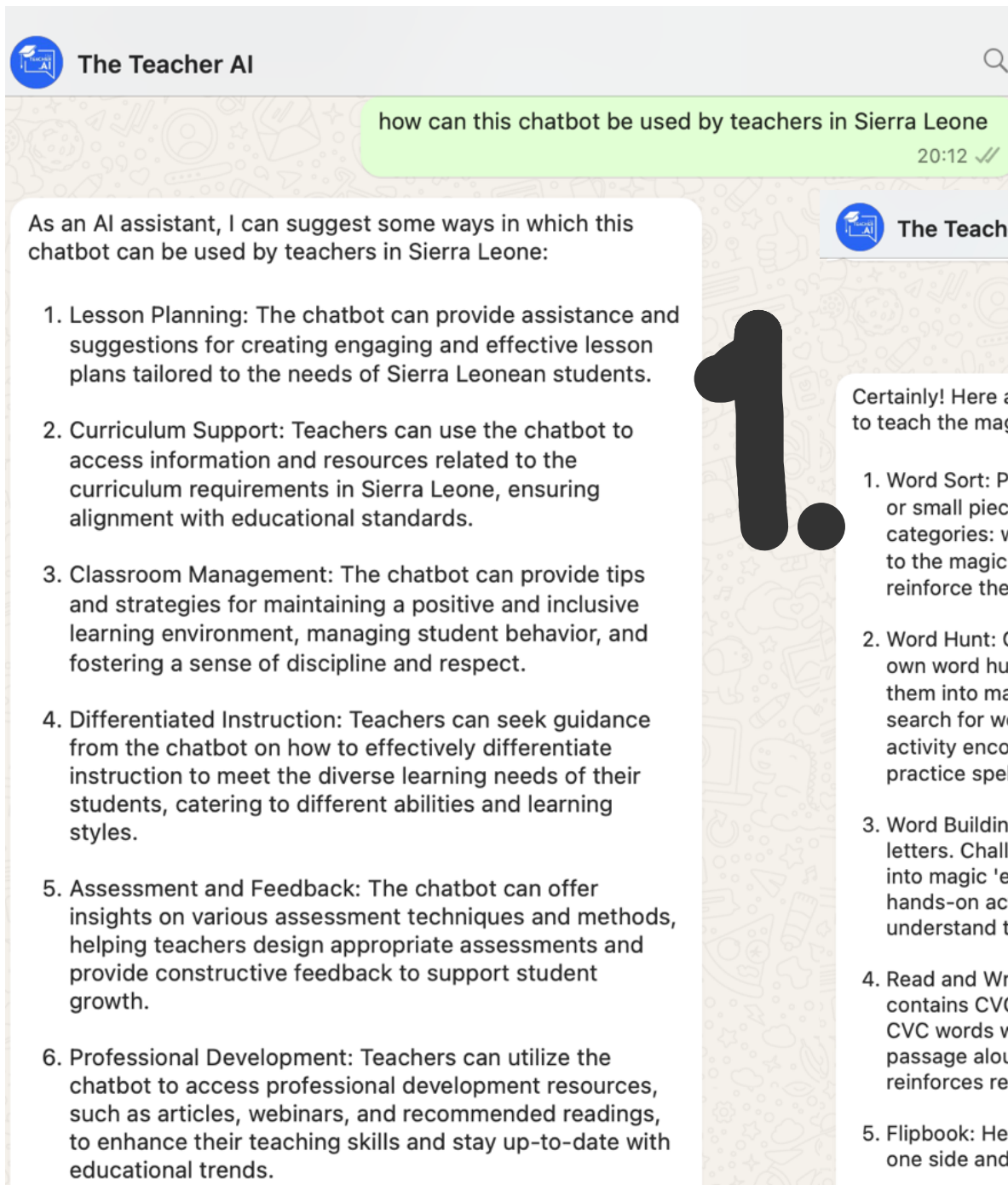
The Teacher AI

how can this chatbot be used by teachers in Sierra Leone
20:12

As an AI assistant, I can suggest some ways in which this chatbot can be used by teachers in Sierra Leone:

1. Lesson Planning: The chatbot can provide assistance and suggestions for creating engaging and effective lesson plans tailored to the needs of Sierra Leonean students.

2. Curriculum Support: Teachers can use the chatbot to access information and resources related to the curriculum requirements in Sierra Leone, ensuring alignment with educational standards.

3. Classroom Management: The chatbot can provide tips and strategies for maintaining a positive and inclusive learning environment, managing student behavior, and fostering a sense of discipline and respect.

4. Differentiated Instruction: Teachers can seek guidance from the chatbot on how to effectively differentiate instruction to meet the diverse learning needs of their students, catering to different abilities and learning styles.

5. Assessment and Feedback: The chatbot can offer insights on various assessment techniques and methods, helping teachers design appropriate assessments and provide constructive feedback to support student growth.

6. Professional Development: Teachers can utilize the chatbot to access professional development resources, such as articles, webinars, and recommended readings, to enhance their teaching skills and stay up-to-date with educational trends.

1.

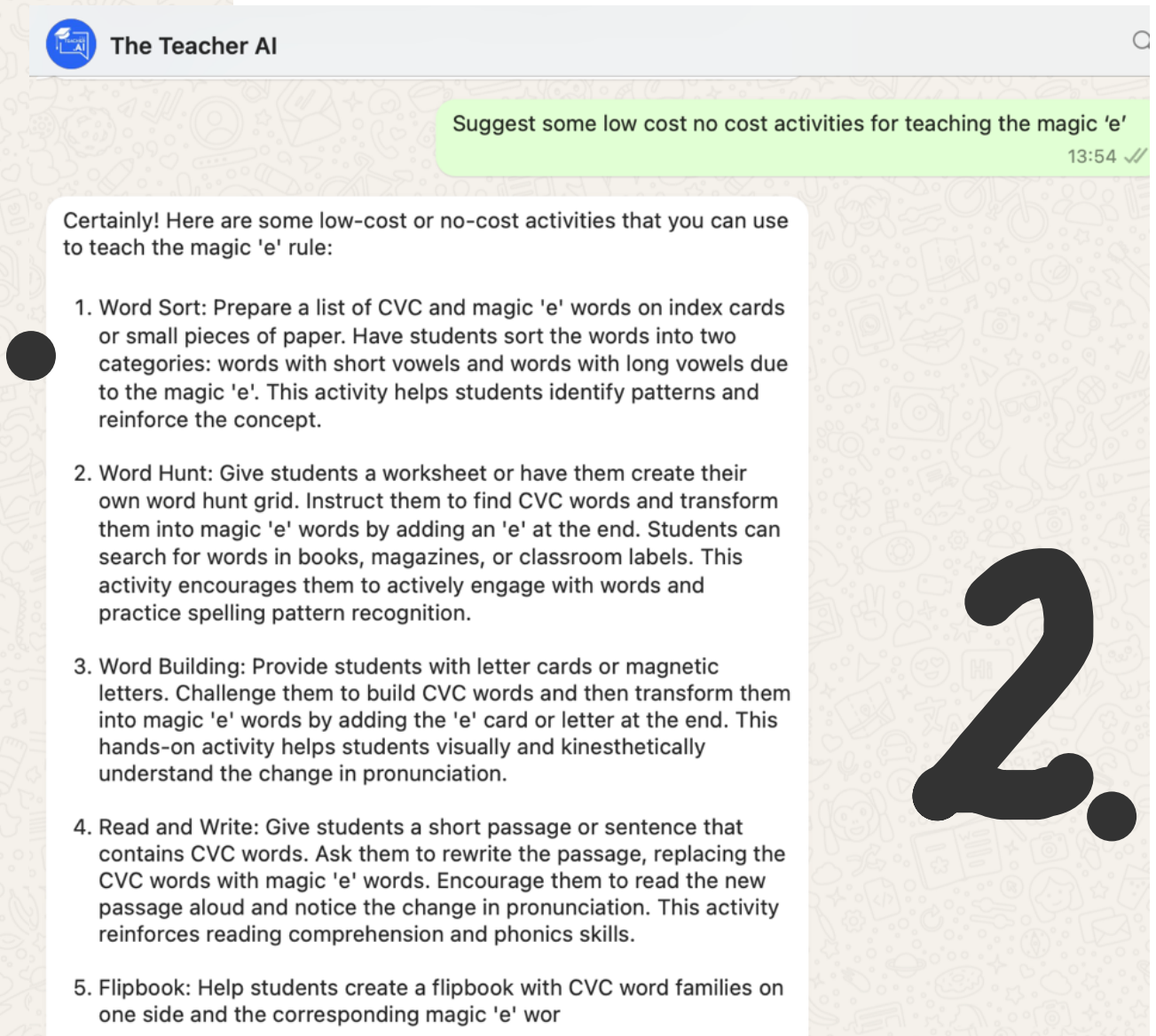
The Teacher AI

Suggest some low cost no cost activities for teaching the magic 'e'
13:54

Certainly! Here are some low-cost or no-cost activities that you can use to teach the magic 'e' rule:

1. Word Sort: Prepare a list of CVC and magic 'e' words on index cards or small pieces of paper. Have students sort the words into two categories: words with short vowels and words with long vowels due to the magic 'e'. This activity helps students identify patterns and reinforce the concept.

2. Word Hunt: Give students a worksheet or have them create their own word hunt grid. Instruct them to find CVC words and transform them into magic 'e' words by adding an 'e' at the end. Students can search for words in books, magazines, or classroom labels. This activity encourages them to actively engage with words and practice spelling pattern recognition.

3. Word Building: Provide students with letter cards or magnetic letters. Challenge them to build CVC words and then transform them into magic 'e' words by adding the 'e' card or letter at the end. This hands-on activity helps students visually and kinesthetically understand the change in pronunciation.

4. Read and Write: Give students a short passage or sentence that contains CVC words. Ask them to rewrite the passage, replacing the CVC words with magic 'e' words. Encourage them to read the new passage aloud and notice the change in pronunciation. This activity reinforces reading comprehension and phonics skills.

5. Flipbook: Help students create a flipbook with CVC word families on one side and the corresponding magic 'e' wor

2.

4. If you wanted to use these ideas for how to teach about the magic 'e' what other questions would you need to ask?

5. What other questions could you ask it about teaching literacy or numeracy? (Try it out.)

n.b. Please note that you have to sense check everything the chatbot sends you. Sometimes it will suggest things that are not real or not good practice. Always sense check everything.

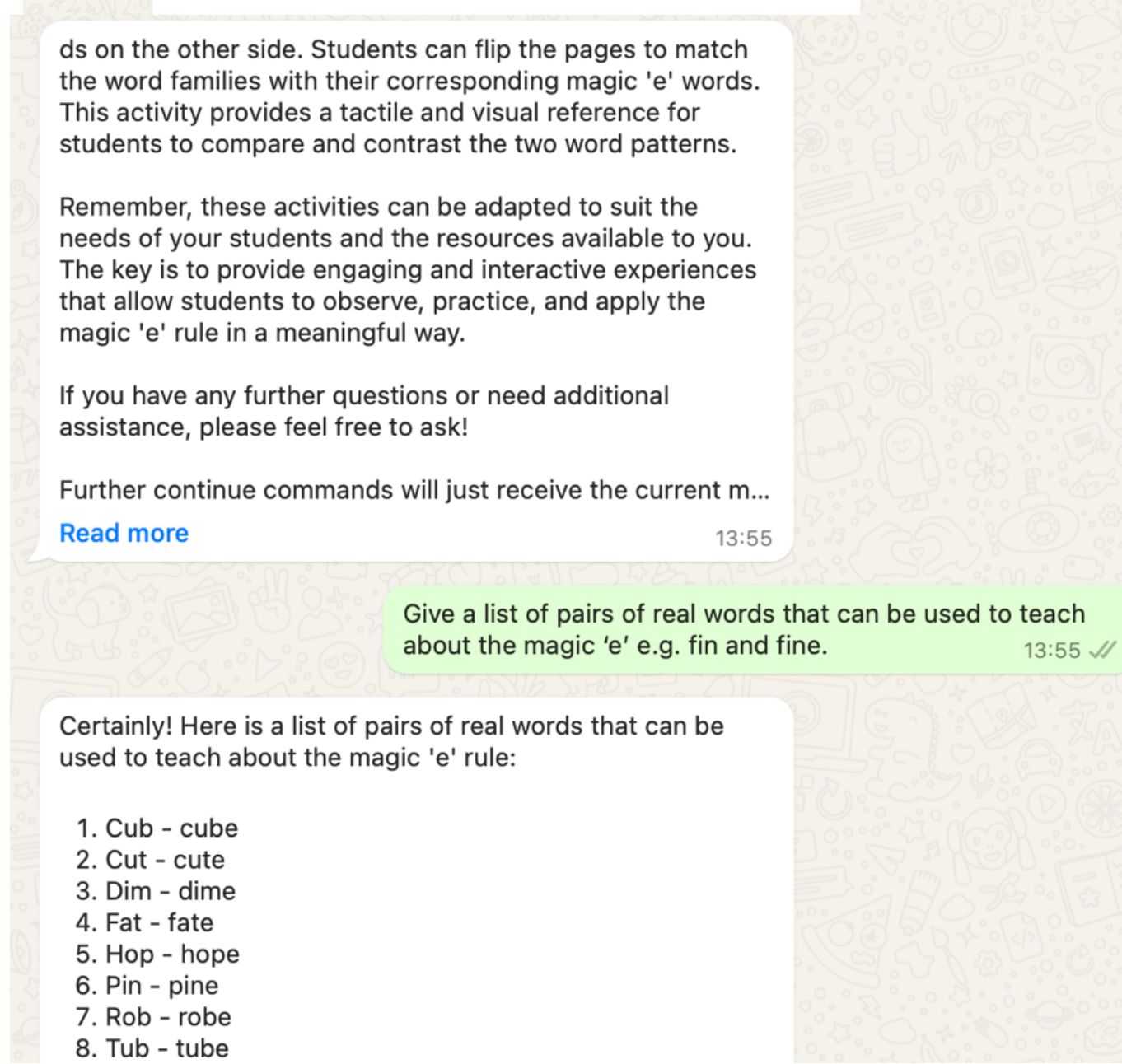
ds on the other side. Students can flip the pages to match the word families with their corresponding magic 'e' words. This activity provides a tactile and visual reference for students to compare and contrast the two word patterns.

Remember, these activities can be adapted to suit the needs of your students and the resources available to you. The key is to provide engaging and interactive experiences that allow students to observe, practice, and apply the magic 'e' rule in a meaningful way.

If you have any further questions or need additional assistance, please feel free to ask!

Further continue commands will just receive the current m...
Read more
13:55

Give a list of pairs of real words that can be used to teach about the magic 'e' e.g. fin and fine.
13:55

Certainly! Here is a list of pairs of real words that can be used to teach about the magic 'e' rule:

1. Cub - cube
2. Cut - cute
3. Dim - dime
4. Fat - fate
5. Hop - hope
6. Pin - pine
7. Rob - robe
8. Tub - tube

**Fig. S9b. Training Brochure.** This is an example of a brochure provided to teachers during onboarding (page 2).

6. Here are five things to bear in mind.

| | |
|---|---|
| **1 Be specific**<br>Clearly state the desired outcome or information needed. This will help the AI to better understand your request and provide more relevant results. | e.g. Don't ask, *'how can I teach English?'* but write: *'give me some activities for teaching sentence formation'* |
| **2 Provide context**<br>Include background information, relevant examples, or any other details that can help the AI understand the context of your query better. | If you want it to use particular vocabulary, or to fit a particular context, you can tell it: e.g., *Write a story using words with the /a/ sound (e.g., apple, ant etc.)* ***or*** *Design activities about good hygiene habits for Sierra Leonean children in class 3.* |
| **3 Use step-by-step instructions**<br>Break down your request into smaller, sequential steps to make it easier for the AI to follow and provide the desired response. | e.g. *Write a (1) 100-word story for (2) class 3 children, about (3) girls' rights, (4) using the words play, day, say, pay and other words that rhyme with them.* |
| **4 Ask for pros and cons**<br>When you need the AI to provide a balanced perspective on a topic or issue, ask it to list the pros and cons. | e.g. *what are the pros and cons of using pair-work activities in class?* |
| **5 Use constraints**<br>To guide the AI's response, set specific parameters, such as word count or the number of bullet points, for the AI to follow. | e.g. *Give* ***5*** *multiple choice questions about flowering plants. Include the correct answer for each one*. ***Or***<br>*Give* ***three*** *activities for practicing complex sentences with 'but'.* |

7. Introduce yourself to TheTeacher.AI so it understands your context. E.g. *I am a class 3 teacher in Sierra Leone. I have 54 children in my class. I have few resources and will need help designing good lessons and activities to teach literacy, numeracy and other subjects.* It will remember things that you told it before, but this will only last for certain number of messages (usually around 10 messages, but it depends on how long the messages and the replies are) so you might have to remind it who you are again later.

8. Try asking TheTeacher.ai the following questions:
   a. I have 3 difficult children in class 4. How can I help them improve their behaviour and attitude?
   b. Use low cost no cost resources. Give me some ideas for how to teach simple punctuation to class 3.
   c. Suggest a 100-word story for 7-year-old children in Sierra Leone using words that rhyme with play.
   d. Give me a simple explanation of the water cycle with 5 low cost no cost activities for how to teach it.
   e. Give the most important 20 words & their definitions for teaching rice-farming to class 5 students.
   f. Explain why corporal punishment will not help improve behaviour.
   g. Suggest some good games for teaching class 3 children how to respect each other.
   h. Suggest a carousel of activities for teaching sentence formation and punctuation.
   i. Suggest a series of 3 activities to engage parents on their responsibilities towards their children's 100% punctual attendance.
9. Which one of these prompts gives you exactly what you want?

10. Which one of these prompts gives answers that require you to ask for adjustments or additional information?

11. Discuss with your group: **3** things you have learned. **2** things that were interesting. **1** remaining question.

**Fig. S9c. Training Brochure.** This is an example of a brochure provided to teachers during onboarding (page 3).

## D. Tables

**Table S1. Domains of search results.** Domains of the top 5 search results for teachers' queries, when submitted to google.com.sl, for 12.5% subsample of teachers' first of thread queries.

| Domain | Proportion |
|---|---|
| en.wikipedia.org | 0.0491 |
| byjus.com | 0.0421 |
| merriam-webster.com | 0.0351 |
| quora.com | 0.0331 |
| study.com | 0.0318 |
| britannica.com | 0.0235 |
| scribd.com | 0.0130 |
| brainly.in | 0.0127 |
| dictionary.cambridge.org | 0.0123 |
| toppr.com | 0.0108 |
| investopedia.com | 0.0107 |
| vocabulary.com | 0.0100 |
| youtube.com | 0.00977 |
| vedantu.com | 0.00962 |
| collinsdictionary.com | 0.00939 |
| mbsse.gov.sl | 0.00841 |
| dictionary.com | 0.00833 |
| mbsseknowledgeplatform.gov.sl | 0.00803 |
| sciencedirect.com | 0.00735 |
| indeed.com | 0.00727 |
| researchgate.net | 0.00644 |
| cuemath.com | 0.00621 |
| grammarly.com | 0.00591 |
| linkedin.com | 0.00538 |
| testbook.com | 0.00500 |
| homework.study.com | 0.00492 |
| brainly.com | 0.00454 |
| my.clevelandclinic.org | 0.00439 |
| pmc.ncbi.nlm.nih.gov | 0.00401 |
| scribbr.com | 0.00401 |
| Other | 0.597 |
| | |
| N | 13,206 |

**Table S2. Query Example: Retrieving Factual Information.** Example query, the response from the AI chatbot, and one of the top 5 search results.

**Query**

"The history of basketball"

| | **AI response** | **Search result** |
|---|---|---|
| | Basketball originated in the United States in the late 19th century. It was invented by Dr. James Naismith, a Canadian physical education instructor, in December 1891. The game was created as a way to keep his students active during the winter months. The first game of basketball was played with a soccer ball and two peach baskets as goals.<br><br>Over time, basketball gained in popularity and became a widely played sport. It was introduced to schools and colleges, and eventually, professional leagues were formed. The National Basketball Association (NBA) is now one of the most popular and influential basketball leagues in the world.<br><br>… | 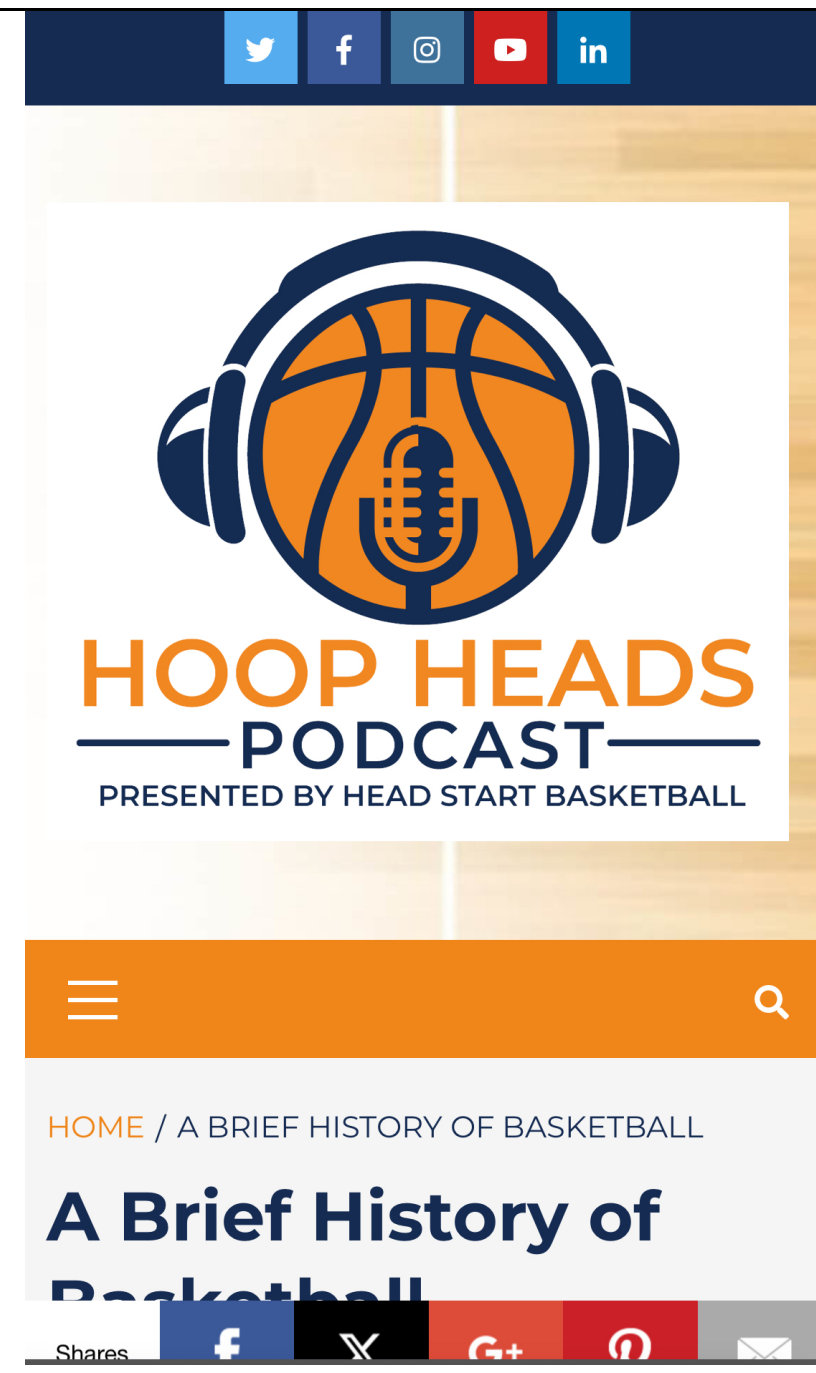<br>https://hoopheadspod.com/a-brief-history-of-basketball/ |
| Word Count | 276 | 1,026 |
| Data Transfer | 1,755 bytes | 9,420,807 bytes |
| Does search result contain main ideas of AI response? | 7.5/10 | |
| Number of Unique Answers | 11-500 answers<br><br>Rationale: "The question about the history of basketball can have many correct answers, focusing on significant events, figures, and developments in the sport's evolution. These responses could be drawn from various texts or resources." | |

**Table S3. Query Example: Lesson Planning.** Example query, the response from the AI chatbot, and one of the top 5 search results.

| | | |
|---|---|---|
| | **Query** | |
| | "How to teach fraction for class six children in Sierra Leone age 11years" | |
| | **AI response** | **Search result** |
| | Teaching fractions to 11-year-old students in Sierra Leone can be made practical and engaging. Here are some tips:<br><br>**Use Visual Aids:** Use visuals like fraction circles or bars to help students understand the concept of parts of a whole.<br><br>**Real-Life Examples:** Relate fractions to everyday situations, like sharing a cake equally among friends, to make it more relatable.<br><br>**Hands-On Activities:** Involve students in hands-on activities like cutting fruits into equal parts to demonstrate fractions practically.<br><br>… | 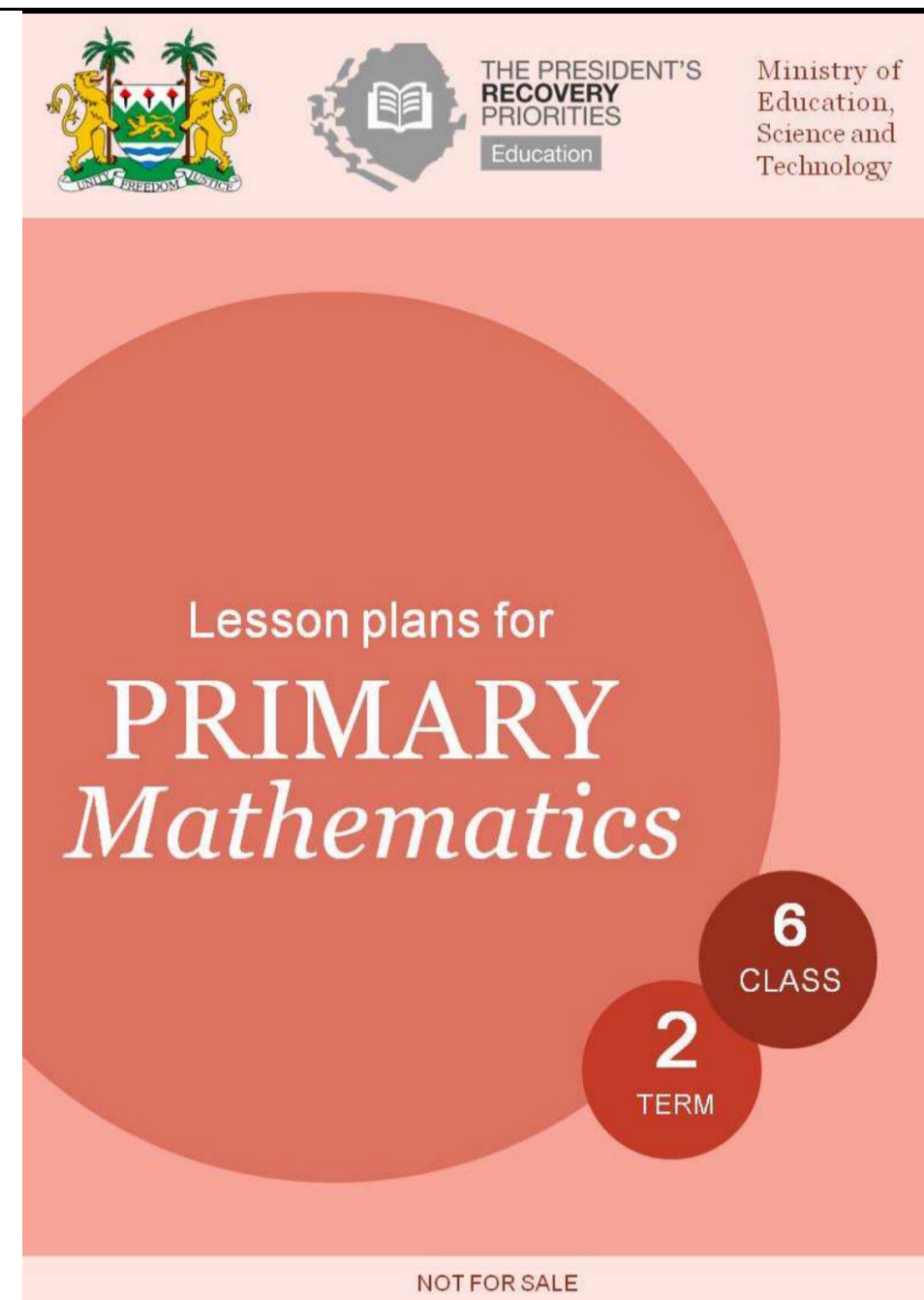<br><br>https://mbsseknowledgeplatform.gov.sl/wp-content/uploads/2022/09/Lesson-Plans-for-Primary-6-Mathematics-Term-2.pdf |
| Word Count | 123 | 50,352 |
| Data Transfer | 844 bytes | 2,234,701 bytes |
| Does search result contain main ideas of AI response? | 5/10 | |
| Number of Unique Answers | 11-500 answers<br><br>Rationale: "The question asks for ways to teach a specific topic (fractions) to a particular age group (11 years old), which could have many correct answers based on different teaching methods, strategies, and educational materials." | |

**Table S4. Query Example: Story.** Example query, the response from the AI chatbot, and one of the top 5 search results.

| | Query | |
|---|---|---|
| | "Please tell us a story for class one kids" | |
| | **AI response** | **Search result** |
| | Once upon a time, there was a little girl named Lily. Lily loved to explore the world around her and learn new things. One day, she went on an exciting adventure to the forest with her friends.<br><br>As they walked through the tall trees, they discovered a hidden treasure chest. They couldn't wait to see what was inside! Lily and her friends carefully opened the chest and found a magical book.<br><br>The book was filled with colorful pictures of animals, plants, and places from all around the world. Lily and her friends were amazed and wanted to learn more about everything they saw in the book.<br><br>They decided to create their own animal encyclopedia. … | 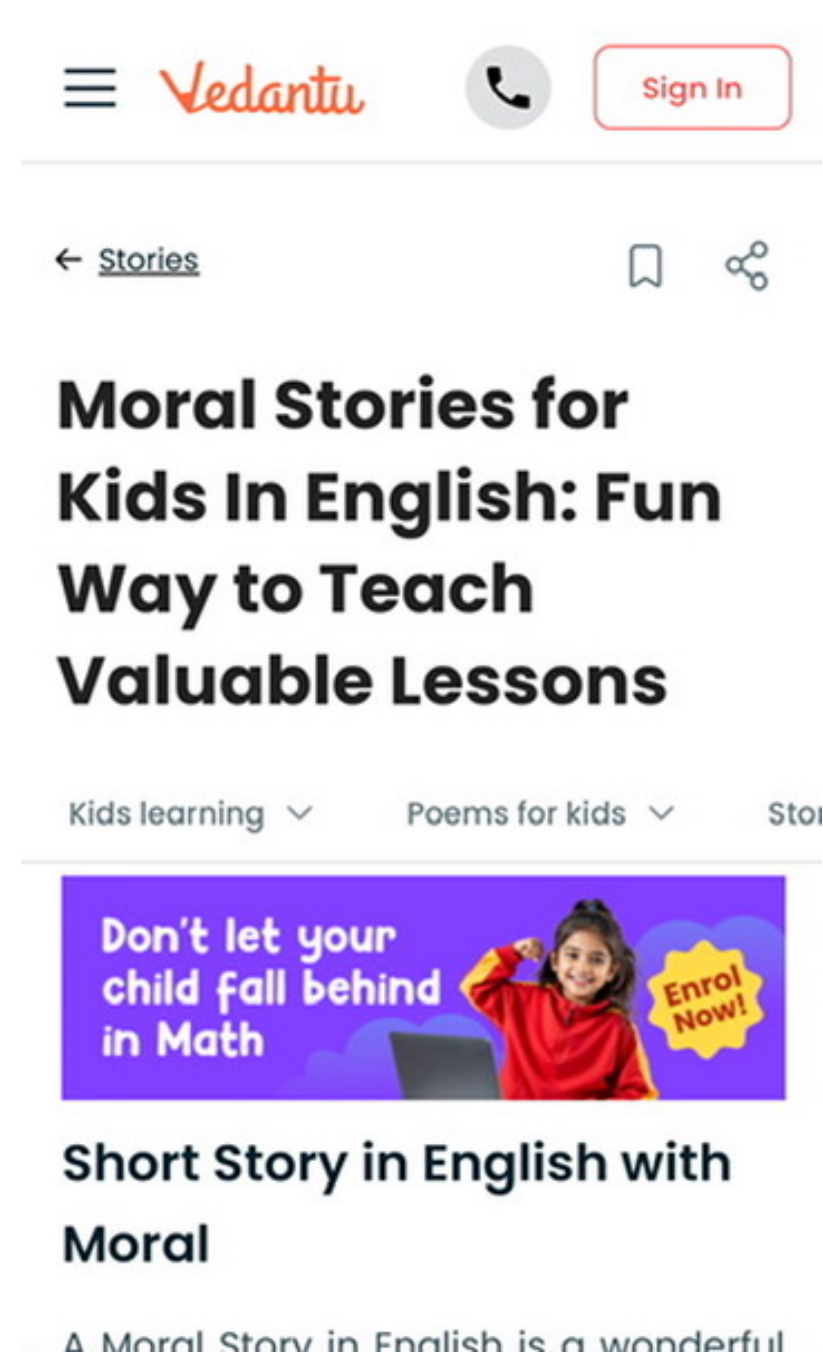<br>https://www.vedantu.com/stories/moral |
| Word Count | 264 (first message) | 1,293 |
| Data Transfer | 1,562 bytes | 3,307,511 bytes |
| Does search result contain main ideas of AI response? | 2/10 | |
| Number of Unique Answers | Infinite<br><br>Rationale: "This query allows for an infinite number of creative responses, as stories can take many different forms and ideas." | |

**Table S5. Comparison of Web Page Data Usage Measurement.** A random sample of 16 search results was manually downloaded with Google Chrome DevTools using settings to mimic a low end mobile device using 3G. Google Chrome downloads some additional components such as dynamic ads; in many cases bandwidth usage continues to grow when the page remains open beyond the period used in our main measure.

| | **Our Main Measure** | **Manual Validation** |
|---|---|---|
| | **Scraping Algorithm** | **Google Chrome DevTools** |
| | Bytes used | Bytes used |
| Mean | 1,185,199 | 5,998,688 |
| SD | 835,896 | 9,414,795 |
| T-test of equality of means (paired, two sided) | p-value: 0.04 | |
| N | 16 | 16 |

**Table S6. Validation of Automated Web Page Text Extraction.**

Difference between manually extracted most relevant text and automated extraction using Trafilatura (percent of total words)

| | 10 percentile | 25 percentile | 50 percentile | 75 percentile | 90 percentile |
|---|---|---|---|---|---|
| Deletions | 0.0 | 0.0 | 0.0 | 1.60 | 6.42 |
| Insertions | 0.0 | 0.04 | 4.08 | 28.13 | 319.90 |

**Table S7. AI Cost Assumptions.** Sources: https://web.archive.org archive of https://openai.com/api/pricing/; https://tech.co/news/chatgpt-openai-update-price-reduction. gpt-oss-20b is an open model which is offered by many different providers; we report the lowest price quote available from https://openrouter.ai/openai/gpt-oss-20b as of 2025-11-24. Each new model is rated more highly than the previous on LMArena.com as of 2025-11-24.

| Model | Release date | Cost per million Input Tokens | Cost per million Output Tokens |
|---|---|---|---|
| GPT 3 Davinci | 2022-09-01 | $20 | $20 |
| GPT 3.5 Turbo | 2023-03-01 | $2 | $2 |
| GPT 3.5 Turbo | 2023-06-14 | $1.50 | $2 |
| GPT 4o mini | 2024-07-18 | $0.15 | $0.60 |
| gpt-oss-20b | 2025-08-05 | $0.04 | $0.04 |

**Table S8. Cost Scenarios.** This table models different cost assumptions, given AI costs in Table S7 and mobile data transfer cost in Sierra Leone of $p_{transfer} = \$1.01$ per gigabyte as of 2024 (*16*). For the cleanest comparison, our main results consider loading a single web page versus AI without a chat history. For search, we present scenarios loading multiple web pages from search, and omitting the download of various components (note that many websites will not function if these components are not loaded). For AI we also present several scenarios. As described in Appendix B1, the implemented chatbot includes the system message both at the beginning and end of the query (average 1377.8 input tokens, 157.1 output tokens); costs are lowered if it is included only once. Alternately, costs increase if we allow the AI to use a chat history filled to 3,596 tokens, the limit at the time. Our AI costs include both the cost of bandwidth and compute costs; the final row displays the bandwidth costs omitting compute costs (the cost faced by teachers in our study, and a lower bound). We compute the number of tokens in text using the OpenAI tiktoken library, using the tokenizer associated with gpt-3.5-turbo. We compute the bytes necessary to transfer a string of text using the UTF-8 encoding.

| | **Cost** ($ per 1000 queries) | | | |
|---|---|---|---|---|
| | Sept 2022 | Sept 2023 | Sept 2024 | Sept 2025 |
| **Search Results** | | | | |
| <u>Standard browser: loading all content (mean unless noted)</u> | | | | |
| Load single web page per query (main estimate) | $2.58 | $2.58 | $2.58 | $2.58 |
| Load 2 web pages per query | $5.17 | $5.17 | $5.17 | $5.17 |
| Load 3 web pages per query | $7.75 | $7.75 | $7.75 | $7.75 |
| Load 4 web pages per query | $10.3 | $10.3 | $10.3 | $10.3 |
| Load 5 web pages per query | $12.9 | $12.9 | $12.9 | $12.9 |
| Load single web page per query (median) | $1.95 | $1.95 | $1.95 | $1.95 |
| <u>Incomplete loading (web page may not function; mean)</u> | | | | |
| Load single web page per query, omit images, video, and audio | $1.77 | $1.77 | $1.77 | $1.77 |
| Load single web page per query, omit all but text/HTML, documents, and code | $1.47 | $1.47 | $1.47 | $1.47 |
| Load single web page per query, omit all but text/HTML and documents | $0.17 | $0.17 | $0.17 | $0.17 |
| Load 2 web pages per query, omit all but text/HTML and documents | $0.34 | $0.34 | $0.34 | $0.34 |
| Load 3 web pages per query, omit all but text/HTML and documents | $0.51 | $0.51 | $0.51 | $0.51 |
| Load 4 web pages per query, omit all but text/HTML and documents | $0.68 | $0.68 | $0.68 | $0.68 |
| Load 5 web pages per query, omit all but text/HTML and documents | $0.85 | $0.85 | $0.85 | $0.85 |
| **AI** | | | | |
| AI response (mean: main estimate)<br>with doubled system prompt and no chat history | $30.7 | $2.38 | $0.30 | $0.06 |
| AI response (median)<br>with doubled system prompt and no chat history | $29.9 | $2.31 | $0.28 | $0.06 |
| AI response (mean)<br>with single system prompt and no chat history | $17.0 | $1.36 | $0.20 | $0.03 |
| AI response with filled chat history and mean output<br>(3,596 input tokens = 4,096-500 tokens) | $75.1 | $5.71 | $0.64 | $0.15 |
| AI response (mean)<br>cost of bandwidth | $0.001 | $0.001 | $0.001 | $0.001 |